\documentstyle[12pt, amsmath, amsfonts, epsfig, amssymb, rotating]{article}
\begin{document}
\def\be{\begin{equation}}
\def\ovx{\vec{x}}
\def\ovy{\vec{y}}
\def\ovr{\vec{r}}
\def\ovn{\vec{n}}
\def\eps{\epsilon}
\def\kt{K_{2}(\tau)}
\def\kdeg{K^{\textrm{deg}}}
\def\rdeg{R_{2}^{\textrm{deg}}}
\def\cala{{\mathcal A}}
\def\caln{{\mathcal N}}
\def\cald{{\mathcal D}}
\def\calc{{\mathcal C}}
\def\calh{{\mathcal H}}
\def\pd{\frac{\pi}{2}}
\def\pt{\frac{\pi}{3}}
\def\pn{\frac{\pi}{n}}
\def\pdn{\frac{\pi}{2n}}
\def\dpn{\frac{2\pi}{n}}
\def\rm{\mathbb{R}}
\def\nm{\mathbb{N}}
\def\cm{\mathbb{C}}
\def\zm{\mathbb{Z}}
\def\hm{\mathbb{H}}
\def\sldr{{SL(2, \rm)}}
\def\sldz{{SL(2, \zm)}}
\def\gi{{\Gamma_{\infty}}}
\def\gn{{\Gamma_{n}}}
\def\vold{\textrm{Vol\ }\cald\ }
\def\im{{\mathcal I}m\ }
\def\sn{\sigma_{n}}
\def\tn{\tau_{n}}
\def\fx{\{x\}}
\def\eplus{\eps_{2} M - \eps_{1} N}
\def\emoins{\eps_{2} M + \eps_{1} N}
\def\epm{\eps_{2} M \pm \eps_{1} N}
\def\dm{\frac{1}{2}}
\def\ba{\bar{\alpha}}
\def\equi{\displaystyle{\ \ \mathop{\sim}_{s\to 1}\ \ }}

\title{\bf Periodic orbits contribution to the 2-point correlation form factor
  for pseudo-integrable systems}
\author{E. Bogomolny, O. Giraud, and C. Schmit\\
 Laboratoire de Physique Th\'eorique et Mod\`eles Statistiques 
 \thanks{Unit\'e Mixte de Recherche de l'Universit\'e Paris XI et du CNRS
   (UMR 8626)}\\
Universit\'e de Paris XI, B\^at. 100\\
91405 Orsay Cedex, France}

\maketitle

\begin{abstract}
  
The  2-point correlation form factor, $K_2(\tau)$, for
small values of $\tau$ is computed analytically for typical examples
of pseudo-integrable systems. This is done by explicit calculation of
periodic orbit contributions in the diagonal approximation.
The following cases are considered: (i) plane billiards
in the form of right triangles with one angle $\pi/n$ and (ii)
rectangular billiards with the Aharonov-Bohm flux line. In the first model,
using the properties of the Veech structure, it is shown that
$K_2(0)=(n+\epsilon(n))/(3(n-2))$ where $\epsilon(n)=0$ for odd $n$,
$\epsilon(n)=2$ for even $n$ not divisible by 3, and $\epsilon(n)=6$ for
even $n$ divisible by 3. For completeness we also recall informally the main
features of the Veech construction. In the second model the answer depends on
arithmetical properties of ratios of  flux line coordinates to the
corresponding sides of the rectangle. When these ratios are
non-commensurable irrational numbers, $K_2(0)=1-3\bar{\alpha}+4\bar{\alpha}^2$
where $\bar{\alpha}$ is the fractional part of the flux through the
rectangle when $0\le \bar{\alpha}\le 1/2$ and it is symmetric with respect
to the line $\bar{\alpha}=1/2$ when $1/2 \le \bar{\alpha}\le 1$.
The comparison of these results with numerical calculations of the form
factor is discussed in detail. The above values of $K_2(0)$ differ from all known
examples of spectral statistics, thus confirming analytically the peculiarities of 
statistical properties of the energy levels in  pseudo-integrable systems.

\end{abstract}

\pagebreak

\section{Introduction}
The statistical properties of quantum systems attracted wide attention  in the
last  years (see e.g. \cite{bohigas}). The investigation of many different 
models had led to a few
accepted conjectures which relate statistical distribution of quantum energy
levels with general properties of corresponding classical motion. For
generic systems these conjectures are the following:  for chaotic systems the
level spacing distribution follows the Random Matrix statistics \cite{mehta},
\cite{bohigasgiannoni}; for integrable systems it follows the Poisson statistics
\cite{berrytaborc}. Both conjectures are
supported by a lot of numerical evidences and by some analytical arguments
\cite{andreev}-\cite{marklof}.

These well-established conjectures are applicable only to completely
chao\-tic or integrable models. But  there are systems which are
neither chaotic nor integrable. Noticeable examples of such systems are  plane 
polygonal billiards with all angles, $\alpha_i$,  commensurable with $\pi$
\begin{equation}
\alpha_{i}=\pi \frac{m_{i}}{ n_{i}},
\end{equation}
where $m_{i}$, $n_{i}$ are co-prime integers. In such  systems all trajectories belong to a 2-dimensional surface of genus
\begin{equation}
g=1+\frac{N}{2}\sum_{i}\frac{m_{i}-1}{n_{i}},
\label{genus}
\end{equation}
where $N$ is the least common factor of the $n_{i}$ (see e.g.
\cite{berryrichens}). The case where all
$m_{i}=1$ corresponds to $g=1$ (i.e. to a torus)  which is integrable.
If  some $m_{i}> 1$ trajectories belong to a higher genus surface and,
consequently, the system is not integrable (at least in the usual sense)
but it is not chaotic either since all trajectories belong to a 2-dimensional
surface and cannot cover a 3-dimensional energy surface ergodically as
is required for chaotic systems. For such reasons these  systems are called 
pseudo-integrable. 

A natural  question appears: what is the spectral statistics of pseudo-integrable
systems? Numerical calculations \cite{berryrichens}-\cite{ugerland}
clearly demonstrated that statistical properties of  such systems differ from
standard examples but have many points in common with the statistics of the
3-dimensional Anderson model at the metal-insulator transition point 
\cite{shklovskii}. The full analytical approach to this question meets with
difficulties related mostly with the existence of quickly growing terms in
the trace formula \cite{bogomolny} which do not permit to use standard
methods.

The main purpose of this paper is to compute
analytically the value of the 2-point correlation form factor, $K_2(\tau)$, in
the limit $\tau\rightarrow 0$ for two examples of pseudo-integrable systems,
(i) a plain billiard in the shape of right triangle with one angle equal $\pi/n$ and
(ii) a rectangular billiards with a Bohm-Aharonov flux line inside. We argue
that in the small-$\tau$ limit the diagonal approximation \cite{berry} 
is valid and the problem reduces to the calculation of distributions
of periodic orbit lengths and areas occupied by periodic orbit families.
Though for general pseudo-integrable systems very little is known on this
subject, triangular billiards in the shape of right triangles with angle
$\pi/n$ belong to the so-called Veech polygons \cite{veech}, \cite{vorobets}
and have a hidden group structure which make possible explicit calculation
of necessary quantities. After the calculations we found a finite value of the
2-point correlation form factor at the origin,  $0<K_2(0)<1$, which is
different from both the Poisson distribution (for which $K_2(0)=1$) and 
the random matrix results (where $K_2(0)=0$). Analogous result
has also been obtained for rectangular billiards with a Bohm-Aharonov flux
line. Non-zero values of the 2-point correlation form factor at the origin
confirm peculiar properties of spectral statistics for pseudo-integrable
systems. We also discuss the comparison of theoretical predictions with the
results of extensive numerical calculations. 

The plan of the paper is the following. In Section \ref{sdiagonal} the
discussions of the trace formula and the diagonal approximation for the
2-point correlation form factor are presented. A brief introduction to the
Veech structure of certain pseudo-integrable billiards is given in Section
\ref{veech}. For clarity we start in Section \ref{square} with a simple
example of square billiards where  ideas and methods can  easily be
illustrated. Needed properties of the modular group and the Eisenstein series
are shortly revised in Sections \ref{modular} and \ref{eisensteinseries}. In
Section \ref{group} the Veech group for $\pi/n$ right triangle is derived
and in Section \ref{periodic} the density of periodic orbits for this
triangle is computed. In Section \ref{calcul} the calculation of 2-point
form factor at the origin is performed and the comparison with the results of
numerical calculations is discussed. Section \ref{flux} is dwelt on the
calculation of the 2-point form factor for a rectangular billiard with a flux
line. As in the previous Sections the main point is the calculation of areas
swept by periodic orbits around the flux line. The result depends on
arithmetical properties of ratios of coordinates of the flux line to the
corresponding rectangular sides. In Section \ref{conclusion} concluding
remarks are presented.

\section{The form factor in the diagonal approximation}\label{sdiagonal}

\subsection{The density of states}

The modern semiclassical approximation of multi-dimensional quantum systems is
based on various types of trace formulas which express quantum density of
states (and other quantities as well) through quantities computed in pure
classical mechanics \cite{bb}, \cite{Gutz}, \cite{berrytabor}. The main step
in deriving trace formulas is the semiclassical approximation for the
(advanced)  Green function
\begin{equation}
G_{+}(\ovx,
\ovy)=\sum_{n}\frac{\overline{\Psi}_{n}(\ovx)\Psi_{n}(\ovy)}{E-E_{n}+i\epsilon},
\end{equation}
where ${E_{n}}$ is the set of energy levels and $\Psi_{n}$ the eigenfunctions
as a sum over classical trajectories with energy $E$ connecting initial
point $\vec{x}$ and final point $\vec{y}$ \cite{bb}, \cite{Gutz}
\begin{equation}
G_+(\vec{x},\vec{y})=\sum_{tr}A_{tr}
\exp (\frac{i}{\hbar}S_{cl}-i\frac{\pi}{2}\nu).
\end{equation}
$S_{cl}$ is  the classical action computed along a trajectory, $A_{tr}$ is a
pre-factor depending on the system considered, and $\nu$ is a phase (the
Maslov index) which, roughly speaking, counts points where simple
semiclassical approximation breaks down.

For 2-dimensional free motion (and for 2-dimensional polygonal billiards) 
the semi-classical approximation for $G$ reads (see e.g. $\cite{Gutz}$)
\be
\label{greenf}
G_{+}(\ovx,\ovy)=\sum_{p}\frac{e^{i k l_{p}-i\pd \nu_{p} -i \frac{3
      \pi}{4}}}{\sqrt{8 \pi k l_{p}}},
\end{equation}
where $l_p$ is the geometrical length of the orbit and $k=\sqrt{E} $ is the wave vector (in the units $\hbar=1$ and $m=1/2$).

The knowledge of the Green function permits to find other quantum
quantities as well. In particular the quantum density of states
\be
d(E)=\sum_{n}\delta(E-E_{n})
\end{equation}
may be written by the means of the advanced Green function as
\be
\label{densite}
d(E)=-\frac{1}{\pi}\ \im\int d \ovx \ G_{+}(\ovx, \ovx).
\end{equation}
The contribution from very short trajectories gives the mean level density,
$\bar{d}$, and the integration over the space selects periodic orbit
contributions \cite{bb}, \cite{Gutz} and determines an oscillating
part of level density, $d^{(osc)}(E)$. For example, the density of states of
an integrable rectangular billiard with sides $a$ and $b$ is
\begin{equation}
d(E)=\bar{d}+d^{(osc)}(E).
\end{equation}
Here the smooth part is
\begin{equation}
\bar{d}=\frac{{\mathcal A}}{4\pi},
\end{equation}
where ${\mathcal A}$ is the area of the rectangle (this formula is valid for all
2-dimensional billiards) and the oscillating part is
\begin{equation}
\label{densitepo}
d_{p.o.}(E)=\sum_{p.p.o.}\sum_{n=1}^{\infty}
\frac{\cala_{p}}{4\pi}\frac{1}{\sqrt{2\pi k n l_{p}}} 
e^{i k n l_{p}-i\frac{\pi}{2}\nu_{p}-i\frac{\pi}{4}} + c.c.
\end{equation}
where
\be
\label{lep}
l_p=\sqrt{(2 M a)^{2}+(2 N b)^{2}}.
\end{equation}
In the rectangular
billiard, the lengths of periodic orbits are 4 times degenerate in the sum
$(\ref{greenf})$ because $(\pm M, \pm N)$ give the same length. When the
integral $(\ref{densite})$ is performed, orbits
$(M,N)$ and $(-M,N)$ are absorbed in the same $\cala_p$. The summation in
$(\ref{densitepo})$ is therefore performed over all primitive periodic orbits of
length $l_p$ with $M \geq 0$ repeated $n$ times (an orbit $(M,N)$ and its
time-reverse companion $(-M,-N)$ are counted as two different orbits). In all integrable
billiards periodic orbits are not isolated but belong to families. $\cala_{p}$
is the area of the pencil of periodic orbits of length $l_{p}$.
For the rectangular billiard $\cala_p=2{\mathcal A}$. 

Pseudo-integrable systems considered in the paper belong to the class of
diffractive systems whose characteristic property is the existence of
singularities which make the classical motion undetermined. Each time a
classical trajectory hits a singularity there is no unique way to continue it.
Quantum mechanics smoothes out these singularities and associates with each
(not too strong) singularity a diffraction coefficient,
$D(\vec{n}, \vec{n}')$, (or scattering amplitude) which defines an amplitude
of scattering on this singularity from the initial direction $\vec{n}$ to the final
direction $\vec{n}'$.

Correspondingly, the semiclassical approximation of the Green function in
the presence of a singularity at point $\vec{x}_0$ takes the form
\be
G(\ovx, \ovy)=G_{0}(\ovx, \ovy)+\sum_{\ovn, \ovn'}G_{0}(\ovx,(\ovx_{0}, \ovn)) 
D(\ovn, \ovn') G_{0}((\ovx_{0}, \ovn'), \ovy),
\end{equation}
where $G_{0}(\ovx, \ovy)$ is the Green function without singularity
and $G_{0}(\ovx, (\ovx_{0}, \ovn))$ is a contribution to the Green function
from a classical trajectory starting at point $\ovx$ and ending at the singularity
$\ovx_{0}$ with momentum in the direction $\ovn$. $G_{0}((\ovx_{0}, \ovn'), \ovx')$
is  a contribution to the Green function from a classical trajectory starting
at point $\ovx_{0}$ with momentum in the direction $\ovn'$ and ending at point
$\ovx'$. 

This modification of the Green function changes the trace
formula. For diffractive systems the density of states can now be written as 
the sum of three terms \cite{keller}, \cite{rosenquist}, \cite{pavloff}
\be
\label{densitetrois}
d(E)=\bar{d}+d_{p.o.}(E)+d_{d.o.}(E),
\end{equation}
where $\bar{d}$ is the mean level density, $d_{p.o.}$ is the contribution of
periodic orbits without singularity, and the third term, $d_{d.o.}(E)$, is a
contribution from all classical orbits starting and ending at the
singularity (with, in general, different momenta). These trajectories are
called diffractive orbits and $d_{d.o.}(E)$ is a sum over all possible
combinations of them
\be
d_{d.o.}(E)=\sum_{m=1}^{\infty}\frac{1}{\pi m}\frac{\partial}{\partial
  E}\sum G(\vec{n}_{1}, \vec{n}_{1}')
D(\vec{n}_{1}', \vec{n}_{2})\ldots
G(\vec{n}_{m-1}, \vec{n}_{m}') D(\vec{n}_{m}', \vec{n}_{1}),
\label{densitedo}
\end{equation}
where $G(\vec{n}, \vec{n}')$ is the contribution to the Green function from
a classical trajectory starting at the singular point with initial momenta
in direction $\vec{n}$ and ending at it with final momentum in direction
$\vec{n}'$.

For polygonal billiards the vertices with $m_i\neq 1$ play the role of 
singular point \cite{keller}. In the case of scattering on the angle $\alpha$ 
the diffraction coefficient can
be derived from Sommerfeld's exact solution \cite{keller}
\begin{equation}
D(\theta_f, \theta_i)=\frac{2}{\gamma}\sin \frac{\pi}{\gamma}\left [
\frac{1}{\cos \pi/\gamma -\cos (\theta_f+\theta_i)/\gamma}-
\frac{1}{\cos \pi/\gamma -\cos (\theta_f-\theta_i)/\gamma}\right ],
\label{dtriangle}
\end{equation}
where $\gamma=\alpha/\pi$ and $\theta_f$ (resp. $\theta_i$) is the final
(resp. initial) scattering angle. 

For rectangular billiards with Aharonov-Bohm flux lines the flux lines 
themselves are
singular points and the exact solution for an infinite plane with a flux
line carrying a flux $\alpha$ \cite{bohmaharonov} gives
\begin{equation}
D(\theta_f, \theta_i)=\frac{2\sin \pi \alpha}{\cos
  \frac{\theta_f-\theta_i}{2}}
e^{i(\theta_f-\theta_i)/2}.
\label{dflux}
\end{equation}
The main difference between pseudo-integrable models discussed in this paper
and  usual diffractive models is the divergence of diffraction coefficients
(\ref{dtriangle}) and (\ref{dflux}) at certain directions (called optical
boundaries because in the simplest case they separate illuminated regions
from dark ones). Of course, exact solutions do not diverge even in vicinity of
optical boundaries. The divergence comes from artificial separation of exact
waves into geometrical and diffraction parts. Nevertheless, this formal
divergence has profound effects on the structure of the trace formula. First,
multiple diffraction along optical boundaries need a special
treatment. Using a kind of uniform approximation in \cite{bogomolny} it was
demonstrated that for polygonal billiards such multiple diffraction produces
terms proportional up to a numerical factor to $l/k$ where $l$ is the total
length of the diffractive orbit. When $l$ is fixed and $k\rightarrow \infty$
(as in the usual approach to trace formulas) these terms are smaller than
periodic orbit terms (\ref{densitepo}) but bigger than diffractive terms
(\ref{densitedo}).
But to compute spectral correlation functions one needs to consider a limit
when $k$ is fixed and $l\rightarrow \infty$. In this limit multiple diffraction
terms are bigger than both periodic orbit and diffraction terms. Another
difficulty is related with the existence of terms corresponding to
diffraction not exactly on optical boundaries but sufficiently close to them
so their contributions are also large. Without exact summation of these
quickly growing terms it is not possible to find spectral statistics of the
systems considered. 

It the next Section we argue that, nevertheless, these terms give negligible
contribution to the value of the 2-point correlation form factor at the
origin and only diagonal contributions of periodic orbits will be important
for this quantity.

\subsection{The 2-point correlation form factor}

The  2-point correlation function is related with the level density by the
formal expression
\be
\label{corr}
R_{2}(\eps)=\left< d(E+\frac{\eps}{2})\ d(E-\frac{\eps}{2}) \right>,
\end{equation}
where the brackets denote an energy averaging around $E$ on an energy window much
larger than the mean level spacing $1/\bar{d}$, and much smaller than 
energy $E$.

The two-point correlation form factor is the Fourier transform of 
$R_{2}(\epsilon)$ :
\be
\label{defk}
K_2(\tau)=\int_{-\infty}^{\infty}\frac{d\eps}{\bar{d}}
\left<d(E+\frac{\eps}{2})\ d(E-\frac{\eps}{2}) \right> e^{2 i \pi \bar{d}
    \eps\tau},
\end{equation}
(the factors are chosen so that $\tau$ and $K_{2}$ are dimensionless). 

Trace formulas, roughly speaking, state that the density of states can be
represented as a sum over classical orbits (both periodic and diffractive)
\begin{equation} 
d^{(osc)}(E)=\sum_{p} C_{p}e^{iS_{p}(E)/\hbar} +c.c.
\end{equation}
Substituting this formal expansion into (\ref{corr}) and using the expansion
$$S(E+\epsilon)\approx S(E)+T(E)\epsilon$$ 
where $T(E)$ is the period of classical motion one obtains \cite{berry}
\begin{eqnarray}
&&\left<d(E+\frac{\eps}{2})\ d(E-\frac{\eps}{2})\right>=\\
&& \sum_{p_1,p_2} C_{p_1}C_{p_2}^* 
<\exp \frac{i}{\hbar}(S_{p_1}(E)-S_{p_2}(E)) >
e^{i(T_{p_1}+T_{p_2})\epsilon/(2\hbar)}.\nonumber
\label{19}
\end{eqnarray}
Here the terms corresponding to the sum of actions are omitted as it is
assumed that they are washed out by the smoothing procedure.

The corresponding expression for the 2-point correlation form factor is the
following:
\begin{equation}
\label{kz}  
K_2(\tau)=\sum_{p_1,p_2} \frac{2\pi \hbar}{\bar{d}}C_{p_1}C_{p_2}^* 
<e^{i(S_{p_1}(E)-S_{p_2}(E))/\hbar} >
\delta (\frac{T_{p_1}+T_{p_2}}{2}-2\pi \hbar \bar{d} \tau).
\end{equation}
The main difficulty in such an approach is the computation of the mean value of 
terms with action differences
\begin{equation}
F(E)=<e^{i(S_{p_1}(E)-S_{p_2}(E))/\hbar} >.
\end{equation}
The best developed approximation (called the diagonal approximation) consists 
in taking into account only terms with exactly the same actions \cite{berry} i.e.
\begin{equation}
F(E)=\left \{ \begin{array}{cc} 1,\;\;\mbox{if}\;S_{p_1}(E)=S_{p_2}(E)\\
0,\;\;\mbox{if}\;S_{p_1}(E)\neq S_{p_2}(E)\end{array}\right . , 
\label{diagonal}
\end{equation}
since terms with $S_{p_1}(E)\neq S_{p_2}(E)$ will vanish by smoothing over $E$.
In this approximation (assuming that for orbits with equal actions
pre-factors are also equal (which is not always the case)) the  2-point 
correlation form factor takes the form
\begin{equation}
K_2^{(diag)}(\tau)=\sum_{p} \frac{2\pi \hbar}{\bar{d}}g_p^2|C_{p}|^2 
\delta (T_{p}-2\pi \hbar \bar{d} \tau),
\label{kdiag}
\end{equation}
where $g_p$ is the multiplicity of a given periodic orbit (i.e. the number
of orbits with exactly the same action) and the summation is performed over
orbits with different actions. In particular for integrable and
pseudo-integrable systems from Eq.~(\ref{densitepo}) one gets
\begin{equation}
K_2^{(diag)}(\tau)=\frac{1}{8\pi^2 \bar{d}} \sum_{p}
\frac{|\cala_p|^2}{\l_p}g_p^2 \delta (l_p-4\pi k \bar{d}\tau ),
\label{kperiodic}
\end{equation}
where as before $l_p$ is the length of a periodic orbit and $\cala_p$ is the
surface occupied by a periodic orbit family. 

It is instructive to perform the calculation for the simplest example
of the rectangular billiard with sizes $a$ and $b$. A periodic orbit in this
billiard is defined by 2 integers $m$, $n$ and its length is 
\be
l_{p}=\sqrt{(2 m a)^{2}+(2 n b)^{2}}.
\end{equation}
As pairs $(m,n)$ and $(m,-n)$ belong to the same family (or torus) the
degeneracy is $g_p = 2$ (we remind the reader that in the rectangular
billiard the terms corresponding to $m<0$ are already taken
into account in $\cala_p$), and it is
sufficient to compute the density of periodic orbits with positive $m,n$
\be
\label{sommerho}
\rho(l)=\sum_{m,n\geq 0}\delta(l-l_{p}).
\end{equation}
Changing the summation over integers $(m,n)$ to the integration
and using the substitution $m=r\cos \phi /(2a)$ and $n=r\sin \phi /(2b)$,
one obtains by integrating over $\phi$ from $0$ to $\pi / 2$
\be
\rho(l)=\frac{\pi l}{8 \cala}.
\label{rho}
\end{equation}
Since all families of periodic orbits in the rectangle cover the same
area $\cala_{p}=2\cala$ and the length multiplicity is $g_p=2$, 
the 2-point correlation form factor for the rectangular billiard in the 
diagonal approximation is 
\be
\kt=\frac{2\cala^{2}}{\pi^{2}\bar{d}}\int_{0}^{\infty}\frac{1}{l}\ 
\delta\left( l-4\pi k \bar{d}\tau\right)\rho(l) d l= 1,
\end{equation}
which is the expected value for the form factor of integrable systems
\cite{berry}. 
The diagonal approximation  (\ref{diagonal}) is known (with physical
accuracy) to be valid
for generic integrable systems \cite{berry} and can be modified \cite{bogomolny2}
to compute mean values of  more than 2 actions in the exponent of (\ref{diagonal}). 

For general systems the validity of the diagonal approximation is restricted only
to small values of $\tau$ \cite{berry}, \cite{varenna} and 
it is usually used to compute the first non-zero term of the
expansion of the 2-point correlation form factor in powers of $\tau$.

For diffractive systems with finite diffraction coefficient one can use
the diagonal approximation for both periodic orbit terms and diffractive
terms. But when the diffraction coefficient diverges in certain directions
these calculations lead to difficulties. For example, multiple diffraction
on optical boundaries corresponding to $n$ repetitions of a primitive
periodic orbit in pseudo-integrable billiard gives the following terms 
\cite{bogomolny}
\begin{equation}
  d_{mult.diff.}(E)=\sum_{l,n}\frac{l}{k}c_n\cos(k n l),
\end{equation}
where $c_n$ are certain numerical coefficients. The attempt to use the
diagonal approximation for these terms  leads to the following result 
\begin{equation}
K_2^{(mult.diff.)} \sim k^2 \tau^3,
\end{equation}
if we take into account that the density of primitive periodic orbits in
pseudo-integrable systems (at least for Veech systems (see the next
Section)) differs only by a numerical factor from Eq.~(\ref{rho}). But this
expression contains powers of  momentum $k$ and when $k\rightarrow \infty$
it cannot be correct. All terms corresponding to diffraction on
or close to optical boundaries give similar quickly growing terms which cannot be
treated separately. Without a  ressumation of these terms the determination of
spectral statistics of such models seems not possible. These
arguments suggest the following scenario.  The 2-point form factor is
a sum of two terms
\begin{equation}
\label{twoterms}
K_2(\tau)=f_1(k^{\alpha}\tau)+f_2(\tau),
\end{equation}
where $\alpha$ is a certain positive quantity. The first function, $f_1(x)$,
describes a result of resummation of quickly growing terms connected with
divergence of the diffraction coefficient and  when $x\rightarrow \infty$
$f_1(x)$ should go quickly to zero. The second function, $f_2(x)$, is a
contribution of diffraction far from optical boundaries and can be computed
similarly to ordinary diffraction \cite{gerland} in perturbation series of $\tau$.
Of course, this is only a plausible conjecture and more detailed
investigation should be done to give credit to it. 

Though the divergence of the diffraction coefficient prevents the calculation
of the 2-point correlation form factor in the full range, one can still use
the trace formula (\ref{densitedo}) to find its behavior at the origin, $\tau=0$.
The main point is that, even when the diffraction coefficient formally
diverges, the exact waves remain finite and using a uniform approximation 
\cite{bogomolny} one can demonstrate that the ratio
\begin{equation}
\frac{D(\vec{n},\vec{n}')}{\sqrt{kl}}
\end{equation}
is bounded for all angles, lengths and momenta. Each term in
the diffractive trace formula (\ref{densitedo}) is a product of certain
number of these ratios and the total period of the corresponding composite orbit
which appears due to the derivative over energy. Therefore it is of order of
$\tau$ multiplied by a constant and in the limit $\tau\rightarrow 0$ all
diffractive terms disappear. Only the periodic orbit contribution
(\ref{densitepo}) remains important at small values of $\tau$. 
From Eq.~(\ref{kperiodic}) one concludes that for pseudo-integrable systems
\begin{equation}
K_2(0)=\lim_{\tau\rightarrow 0}\frac{1}{8\pi^2\bar{d}}
\sum_{p} \frac{|\cala_{p}|^2}{l_p}g_p^2 
\delta (l_{p}-4\pi k \bar{d} \tau).
\label{k0}
\end{equation}
The main problem now is how to compute the density of periodic orbits and the
distribution of the areas of periodic orbit families. For
generic pseudo-integrable systems very little is known and no reliable
calculations can be done. E.g. for general
plane polygonal billiards with angles commensurable with $\pi$
it has only been proved $\cite{masurun}, \cite{masurdeux}$ that the number
of periodic orbits with length less than $l$, $\caln (l_p \leq l)$ obeys inequalities
\begin{equation}
c_{1} l^{2}<\caln (l_p \leq l)<c_{2} l^{2}
\end{equation}
for certain  constants $c_{1}$ and $c_{2}$ (depending on the polygon). But
even the existence of an asymptotic law for $\caln (l_p \leq l)$ was not proved.

Fortunately, there is a sub-class of pseudo-integrable billiards for which
all necessary quantities can be computed 
due to the existence of a hidden group structure, 
and the triangular billiard in the shape of the right
triangle with one angle equal to $\pi/n$ belongs to this class. In the following
Section we focus on those polygons.

\section{Veech structures for polygonal billiards}\label{veech}

We start the discussion of a hidden group structure of  certain polygonal billiards
with the simple  example of the square billiard where the necessary
ideas and methods can be illustrated clearly without technical difficulties.

\subsection{A simple case: the square billiard}\label{square}

How can one evaluate the number of periodic orbits with length less than $l$
in a square billiard of size $2a$ with periodic boundary conditions? The 
exact expression for the
length of the periodic orbits in the such  billiard is, of course,
\be
l_{p}=\sqrt{(2 m a)^{2}+(2 n a)^{2}}
\label{pol}
\end{equation}
with $m \in \nm$ and  $n  \in \zm$ (see ($\ref{lep}$)).
The  number of periodic orbits with length less than $l$, 
$\caln (l_{p}\leq l)$, reads
\be
\caln (l_{p}\leq l)=\sum_{m,n}\Theta \left(l-2 a \sqrt{m^{2}+n^{2}}\right)
\end{equation}
and asymptotically when $l \rightarrow\infty$
\begin{eqnarray}
\label{no}  
\caln (l_{p}\leq l)&=&\int_{0}^{\infty} d m \int_{-\infty}^{\infty} d n 
\ \Theta \left(l-2a\sqrt{m^{2}+n^{2}}\right)\nonumber \\
&=&\frac{\pi l^{2}}{8 a^{2}}
\end{eqnarray}
if one sets $m=(r \cos \varphi)/2a$ and  $n=(r \sin \varphi)/2a$. This is
the number of all periodic orbits. 
More interesting questions and rich mathematical structure appear
when one is interested in the calculation of the number of primitive periodic
orbits $\caln_{pp} (l_{p}\leq l)$ (that is, orbits with $m$ and $n$ coprime).

The number of such orbits for a square billiard can easily be computed by
using the inclusion-exclusion principle. The number of primitive periodic
orbits with length less than $l$ is  
the total number of periodic orbits with length less than $l$ minus
the number of orbits repeated $p$ times with prime $p$, to which we add
orbits repeated $p_1p_2$ times, which had been subtracted twice, etc. Finally
one concludes that
\begin{eqnarray}
\caln_{pp} (l_{p}\leq l)&=&\caln (l_{p}\leq l)-
\sum_p\caln (l_{p}\leq \frac{l}{p})+
\sum_{p_1,p_2}\caln (l_{p}\leq \frac{l}{p_1p_2})\nonumber\\
&-&\sum_{p_1,p_2,p_3} \caln (l_{p}\leq \frac{l}{p_1p_2p_3}) \ldots
\end{eqnarray}
Using the $l^{2}$ dependence of $\caln$ in $(\ref{no})$, we have
\begin{eqnarray}
\caln_{pp} (l_{p}\leq l)&=&\caln (l_{p}\leq l)
(1-\sum_p\frac{1}{p^2}+\sum_{p_1,p_2}\frac{1}{(p_1p_2)^2}-
\sum_{p_1,p_2,p_3}\frac{1}{(p_1p_2p_3)^2}\ldots)\nonumber\\
&=&
\caln\prod_p(1-\frac{1}{p^2})=\caln\frac{1}{\zeta (2)}=
\frac{6}{\pi^2}\caln,
\end{eqnarray}
where
\begin{equation}
\zeta (s)=\sum_{n=1}^{\infty}\frac{1}{n^s}=\prod_p\frac{1}{1-p^{-s}}
\end{equation} 
is the Riemann zeta function.

From $(\ref{no})$ one gets
\be
\caln_{pp} (l_{p}\leq l)=\frac{3 l^{2}}{4 \pi a^{2}}.
\label{densiterect}
\end{equation}
Our aim is to generalize the previous calculation of $\caln_{pp} (l_{p}\leq l)$
to certain triangular billiards. This  generalization naturally appears
\cite{veech} when
one considers carefully the usual geometrical picture of the free motion
inside the square billiard. It is well known that any trajectory of such a motion
can be unfolded to a straight line when instead of the square billiard
one considers the motion on the covering space which for square billiard is
a plane with infinite square lattice of the side $2a$. The vertices of this
lattice (which are the images of the vertices of the initial square) have
coordinates
\begin{equation}
x=2a m,\;\; y=2a n
\end{equation}
with integers $m$ and $n$ and can be considered as the result of the
application of a
$2\times 2$ matrix with integer coefficient to a horizontal vector $(2a, 0)$  
\be
\label{matorb}
\left(\begin{array}{cc}m&k \\ n&l\end{array}\right)
\left(\begin{array}{c}2a \\ 0\end{array}\right)=
\left(\begin{array}{c}2 a m \\ 2 a n\end{array}\right).
\end{equation}
Thus, the periodic orbit lengths (\ref{pol}) are the distances between these
vertices and the initial point $(0,0)$.  

The problem of finding the number $\caln_{pp}$ of primitive periodic
orbits with length less than $l$ is therefore equivalent to the problem of
finding out how many
$2\times 2$ matrices with integer coefficients and determinant equal to $1$
(since $m$ and $n$ are coprime one can  impose $m l-n k=1$) exist with
$m^{2}+n^{2}\leq x^{2}$ for a given $x$ (or, which is equivalent, with
$n^{2}+l^{2}\leq x^{2}$). The group of $2\times 2$ matrices with integer
coefficients and determinant equal to $1$ form a group $\sldz$ and 
in the next two Sections we shall discuss its main properties. Though this
material is well known  we find it useful to remind it informally.

\subsection{The modular group}\label{modular}

The subgroup of $\sldr$ containing all $2\times 2$ matrices with integer
coefficients and determinant equal to $1$ is called the modular group 
$\sldz$. The standard representation of this group
(see e.g. $\cite{courburneg})$ is the Poincar\'e half-plane $\calh$ with measure
\be
\label{mesure}
ds^{2}=\frac{1}{y^{2}}(dx^{2}+dy^{2})\ ;
\end{equation}
A matrix $g\in \sldz$ is represented by the isometry
\be
\label{gdz}
\begin{array}{rrcl}
g:&\calh &\rightarrow& \calh \\  
 & z &\mapsto & \displaystyle{\frac{m z+k}{n z+l}}
\end{array}  
\end{equation}
The modular group is generated by the translation $T: z\mapsto z+1$ and
the inversion $S: z\mapsto -1/z$, which correspond respectively to the
matrices 
\be
\left(\begin{array}{cc}1&\alpha \\ 0&1\end{array}\right)
\end{equation}
(with $\alpha=1$ for the modular group) and
\be
\left(\begin{array}{cc}0&1 \\ -1&0\end{array}\right).
\end{equation}
Since the modular group is a discrete group, we can define its fundamental
domain $\cald$ (shown in fig. $\ref{domfondsldz}$), that is the domain of the
Poincar\'e half-plane $\calh$ that covers $\calh$ under the action of the
representation $(\ref{gdz})$ of the group.

\begin{figure}[ht]
\begin{center}
\epsfig{file=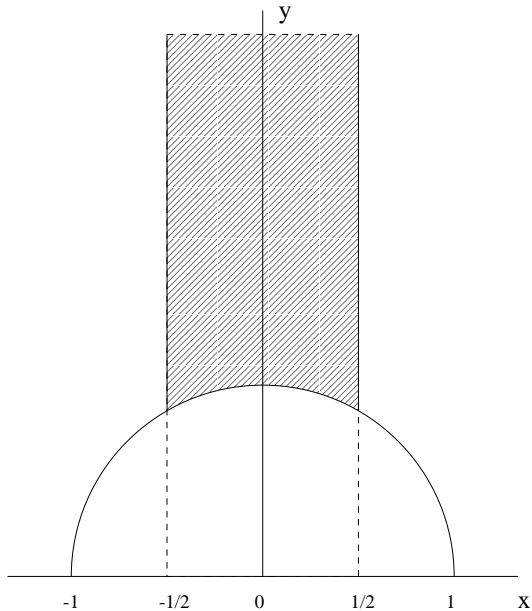,width=7cm}
\end{center}
\caption{The fundamental domain of the modular group}
\label{domfondsldz}
\end{figure}
In order to compute the number of matrices 
$g=\left(\begin{array}{cc}m&k \\n&l\end{array}\right)\in \sldz$ 
verifying $n^{2}+l^{2}\leq x^{2}$, we have to evaluate 
\be
\caln (x)=\sum_{\begin{array}{c}g\in \gi \backslash G \\ n^{2}+l^{2}\leq x^{2}\end{array}} 1
\end{equation}
where  $G=\sldz$ and  $\gi$ is the subgroup of $G$ generated by the translations
($\gi=\{T^{n}, n\in \zm\})$: since the left multiplication by matrices of
the form
$T^p$ 
\begin{equation}
\left(\begin{array}{cc}1&p \\ 0&1\end{array}\right)
\left(\begin{array}{cc} m&k \\ n&l\end{array}\right)=
\left(\begin{array}{cc}m+pn&k+pl\\ n&l\end{array}\right)
\end{equation}
does not change $n$ and $l$ it is necessary to consider
the quotient $\gi \backslash G$ (i.e. 2 matrices which differ by $T^p$ are
considered only once), so that the
sum is convergent $\cite{terras}$. If we assume that in the limit
$x\rightarrow \infty$ the sums can be written as integrals over
$n$ and $l$ with uniform measure (see later) in the form
$(B/\pi)d n d l$, we get
\be
\caln (x)=\int_{n^{2}+l^{2}\leq x^{2}}\frac{B}{\pi}d n d l=\frac{B}{2}x^{2}.
\label{nbinfx}
\end{equation}

\subsection{Eisenstein series}\label{eisensteinseries}

In order to compute the coefficient $B$, let us introduce the Eisenstein series
\be
\label{eisenstein}
E(z, s)=\sum_{g\in \gi \backslash G}\left( \im g(z)\right)^{s}
\end{equation}
for $s>1$. From expression $(\ref{gdz})$ we get, since $m l-n k=1$,
\be
\im g(z)=\frac{y}{|n z+l|^{2}}
\end{equation}
where $y=\im z$. Since $\im g'g(z)=\im g(z)$ for $g'\in \gi$, the sum over
$\gi \backslash G$ is well defined. Let us first compute the asymptotic behavior of
$E(z,s)$ when $s\rightarrow 1$. For a given $R\in \rm$, we can rewrite the sum
$(\ref{eisenstein})$ as a finite sum over elements of $G$ for which
$n^{2}+l^{2}<R^{2}$ and a sum over elements for which $n^{2}+l^{2}>R^{2}$
which diverges as $s \to 1$.
The divergent part, $n^{2}+l^{2}>R^{2}$, is 
\begin{eqnarray}
E^{\textrm{div}}(z,s)&=&\frac{y^{s}}{\pi}\int_{n^{2}+l^{2}>R^{2}}\frac{B\ dn\ dl}{|n
  z+l|^{2s}}\\
&=&\frac{B y^{s}}{\pi}\int_{R}^{\infty}\int_{0}^{\pi}\frac{r^{1-2s}\ dr
  d\phi}{\left[(x \sin\phi+\cos\phi)^{2}+(y \sin\phi)^{2}\right]^{s}}
\end{eqnarray}
Since
\be
\int_{R}^{\infty} r^{1-2s} dr=\frac{R^{2(1-s)}}{2(s-1)}\equi\frac{1}{2(s-1)},
\end{equation}
and the finite part of the Eisenstein series can be neglected as compared
with the divergent part, we have
\be
E(z,s) \equi \frac{y B}{2\pi (s-1)}\int_{0}^{\pi}\frac{d\phi}{(x
  \sin\phi+\cos\phi)^{2}+(y \sin\phi)^{2}}.
\end{equation}
The computation of the integral can be performed the following way : setting
$A=1-x^{2}-y^{2} \ $, $B=2x \ $ and $C=1+x^{2}+y^{2} \ $, we get
\be
\int_{0}^{\pi}\frac{d\phi}{(x^{2}+y^{2})
  \sin^{2}\phi+\cos^{2}\phi+2 x
  \sin\phi\cos\phi}=2\int_{0}^{\pi}\frac{d\phi}{A \cos 2\phi+B\sin 2\phi+C}
\end{equation}
and since $C^{2}-A^{2}-B^{2}=4 y^{2}$, the integral is equal to
\be
\int_{0}^{2\pi}\frac{d\Psi}{\sqrt{A^{2}+B^{2}}\cos\Psi+C}=\frac{\pi}{y}
\end{equation}
Finally
\be
\label{eque}
E(z,s)\equi \frac{B}{2 (s-1)}
\end{equation}
and this limit does not depend on $z$. 

Now let us integrate the series in $(\ref{eisenstein})$ with the invariant
measure $d\mu(z)=d x d y /y^{2}$ over a part, $\cald_{Y}$, of the
fundamental domain $\cald$ corresponding  to a restriction $y\leq Y$. If $d$
is the width of the fundamental domain,
\begin{eqnarray}
\int_{\cald_{Y}}E(z, s)\ d\mu(z)&=&
\sum_{g\in \gi \backslash G} \int_{\cald_{Y}}\left(\im g(z)\right)^{s}\ d\mu(z)
\nonumber\\
&=&\sum_{g\in \gi \backslash G} \int_{g\cald_{Y}}\left(\im z\right)^{s}\ d\mu(z)
\label{inta}\\
&\simeq &\int_{0}^{d}dx\int_{0}^{Y}\frac{dy}{y^{2}}y^{s} 
\label{intb}\\
&=&d \frac{Y^{s-1}}{s-1}.
\label{volume}
\end{eqnarray}
In transformation from Eq.~(\ref{inta}) to Eq.~(\ref{intb}) we take into
account that the image of fundamental region $\cald$ (and $\cald_{Y}$)
under the action of $G$ is a certain region on the upper-half plane which
under the action of $\gi$ can be moved into a vertical strip of width $d$
(which is the fundamental region for the group $\gi$). These images can not
intersect and when $Y\rightarrow \infty$ will cover the whole strip with 
$y\leq Y$.

The asymptotic behavior of this integral is thus the following
\begin{equation}
\lim_{s\rightarrow 1}\int_{\cald_{Y}}E(z, s)\ d\mu(z)= \frac{d}{s-1}.
\end{equation}
By comparing this expression with Eq.~(\ref{eque}) one concludes that
 the value of the constant $B$ is 
\begin{equation} 
B=2 \frac{d}{\vold}
\end{equation}
and from Eq.~(\ref{nbinfx}), the final answer for the density of primitive
($n$ and $l$ are coprime) periodic orbits in a square billiard is
\be
\label{calnveech}
\caln_{pp} (n^{2}+l^{2}\leq x^{2})=\frac{d}{\vold} x^{2}
\end{equation}
For the modular group $d=1$, $\vold=\pi/3$, and for a square billiard with
side $2a$ we have $x=l/(2a)$ according to $(\ref{pol})$, so
\be
\label{calnpp}
\caln_{pp}(l_{p}<l)=\frac{3 l^{2}}{4\pi a^2}
\end{equation}
which agree with Eq.~(\ref{densiterect}) obtained by a different method.

\subsection{Veech group for $\pi / n$ right triangles}\label{group}

\subsubsection{The symmetry group}

The previous calculations were possible because we have found a group
--the modular group-- that relates periodic orbits in a
square to a simple vector (see Eq.~(\ref{matorb})). In order to generalize
this construction for more
complicated polygons it is important to point out that the modular group
is the symmetry group of the unfolding of the square billiard (that is, the lattice
whose unit cell is a $1\times 1$ square). Indeed this square lattice is,
evidently, invariant under the following two transformations:  the rotation
by $\pi/2$ around the center of the square which we denote by $S$  and 
the translation of one coordinate (say $x$) by $1$ which we denote by $T$. 
In Cartesian
coordinates these transformations are represented by the following matrices
\begin{equation}
S=\left (\begin{array}{cc}0&1\\-1&0\end{array}\right )\;\;
T=\left (\begin{array}{cc}1&1\\0&1\end{array}\right ).
\end{equation}
The vertices of the lattice are also unchanged under the action of the group
generated by $S$ and $T$.
But it is well known that this group is exactly the modular group $\sldz$. Therefore
the modular group plays a double role for a square billiard. First, it is the
group of invariance of unfolding of the square and, second, it generates
periodic orbits starting from a fix vector as in Eq.~(\ref{matorb}).

It has been proved by Veech \cite{veech} that for certain polygons (called the Veech
polygons) there exists a group with similar properties. In particular,
a $\pi / n$ right triangle (i.e. a triangle with angles
$\pi/2, \pi/n, \pi (n-2)/2n$) belongs to the Veech polygons \cite{veech},
\cite{vorobets}. 

Let us consider this case in details. The geometrical construction of the
unfolding of the classical trajectories in such a billiard is slightly
different for $n$ even and odd. By reflections with respect to the sides
corresponding to the $\pi/n$
angle the $\pi/n$ triangle can be unfolded to the regular $n$-gon. For $n$
even the opposite sides of this $n$-gon should be identified by translations
(see Fig.~\ref{unfolding}a ). For $n$ odd one has to consider 2 regular
$n$-gons reflected with respect to
one side and to identify parallel sides by translations as in
Fig.~\ref{unfolding}b. The resulting surface is the surface of
genus $(n-1)/2$ for $n$ odd (see \ref{genus}) to which all trajectories belong. 
\begin{figure}[ht]
\begin{center}
\epsfig{file=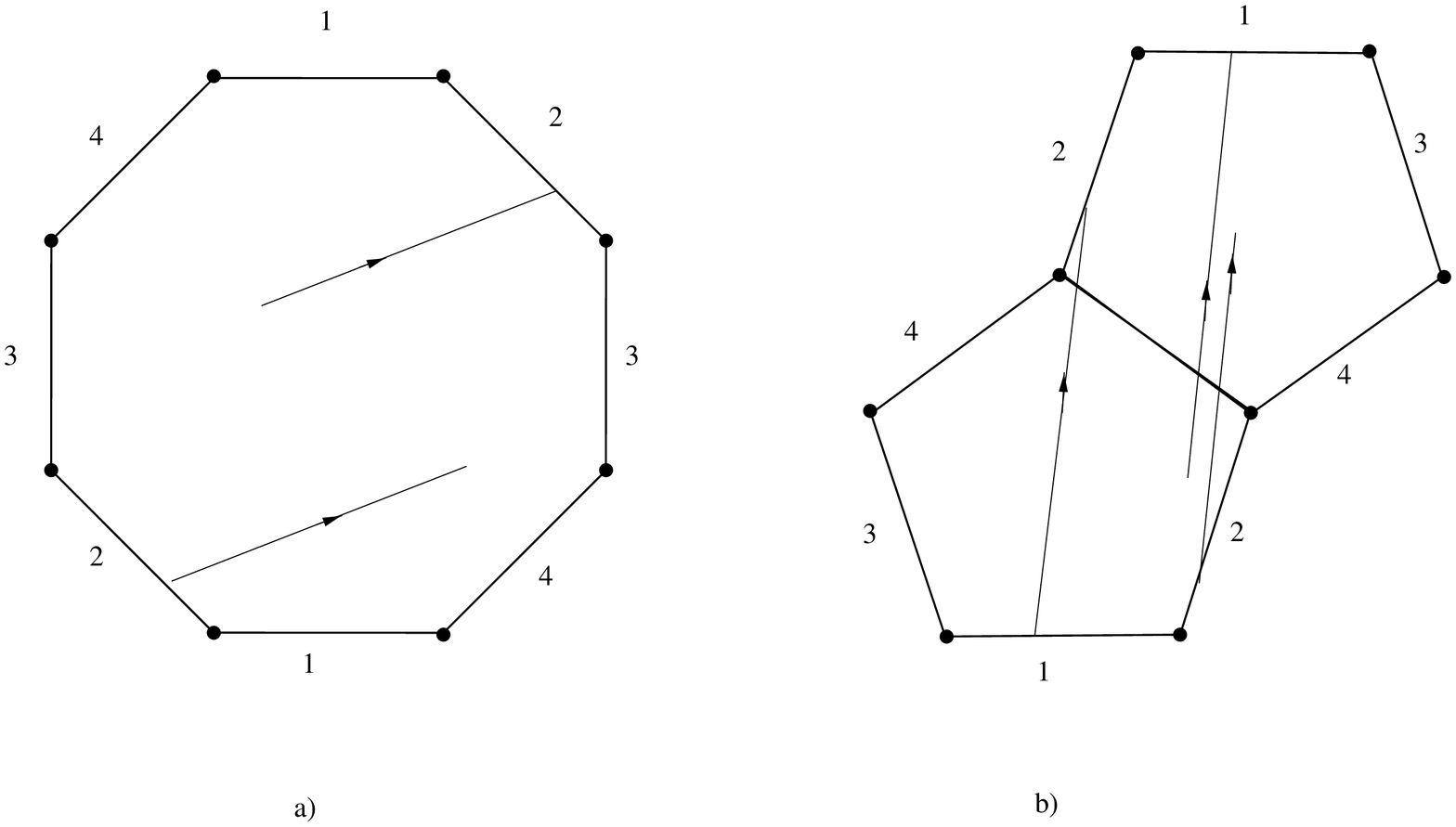,width=13cm}
\end{center}
\caption{The unfolding of $\pi/n$ triangle. Left -- $n$ is even. Right --
  $n$ is odd}
\label{unfolding}
\end{figure}
From this construction it is clear that if a group of invariance exists it
should include the rotation by $2\pi /n$ around the center of these $n$-gons.
In Cartesian coordinates this rotation is defined by the following matrix
\begin{equation}
\sn=\left(\begin{array}{cc}\cos\dpn &-\sin\dpn \\
\sin\dpn&\cos\dpn\end{array}\right).
\label{sigman}
\end{equation}
To find other transformations which leave this surface invariant it is
necessary to consider a few families of periodic orbits.

For $n$ even, we define in the $n-$gon two important elementary
families of periodic orbits: the
first one is the family of horizontal primitive periodic orbits, the second one
is the family of primitive periodic orbits making an angle $\pi/n$ with the
horizontal (see fig. $\ref{octogone}$). For $n$ odd we only define the first
family (see fig. $\ref{pentagone}$). 

For $n$ even ($n=4p$ or $n=4p+2$), the first family has orbits with lengths
\be
\label{longa}
L_{j}=4\cos\frac{(2j-1)\pi}{n}\cos\pn
\end{equation}
and widths
\be
\label{larga}
W_{j}=2 \cos\frac{(2j-1)\pi}{n}\sin\pn
\end{equation}
with $1\leq j \leq p$.
The second  family has orbits with lengths
\be
\label{longb}
l'_{j}=4\cos\frac{(2j-2)\pi}{n}\cos\pn
\end{equation}
and widths
\be
\label{largb}
w'_{j}=2 \cos\frac{(2j-2)\pi}{n}\sin\pn
\end{equation}
with $2\leq j \leq p$ if $n=4p$ or  $2\leq j \leq p+1$ if $n=4p+2$.
The orbit with $j=1$ is special: it has a length and a width equal to
\begin{equation}
l'_{1}=2\cos\pi/n, \;w'_{1}=2\sin\pi/n
\label{exceptional}
\end{equation}

\begin{figure}[ht]
\begin{center}
\epsfig{file=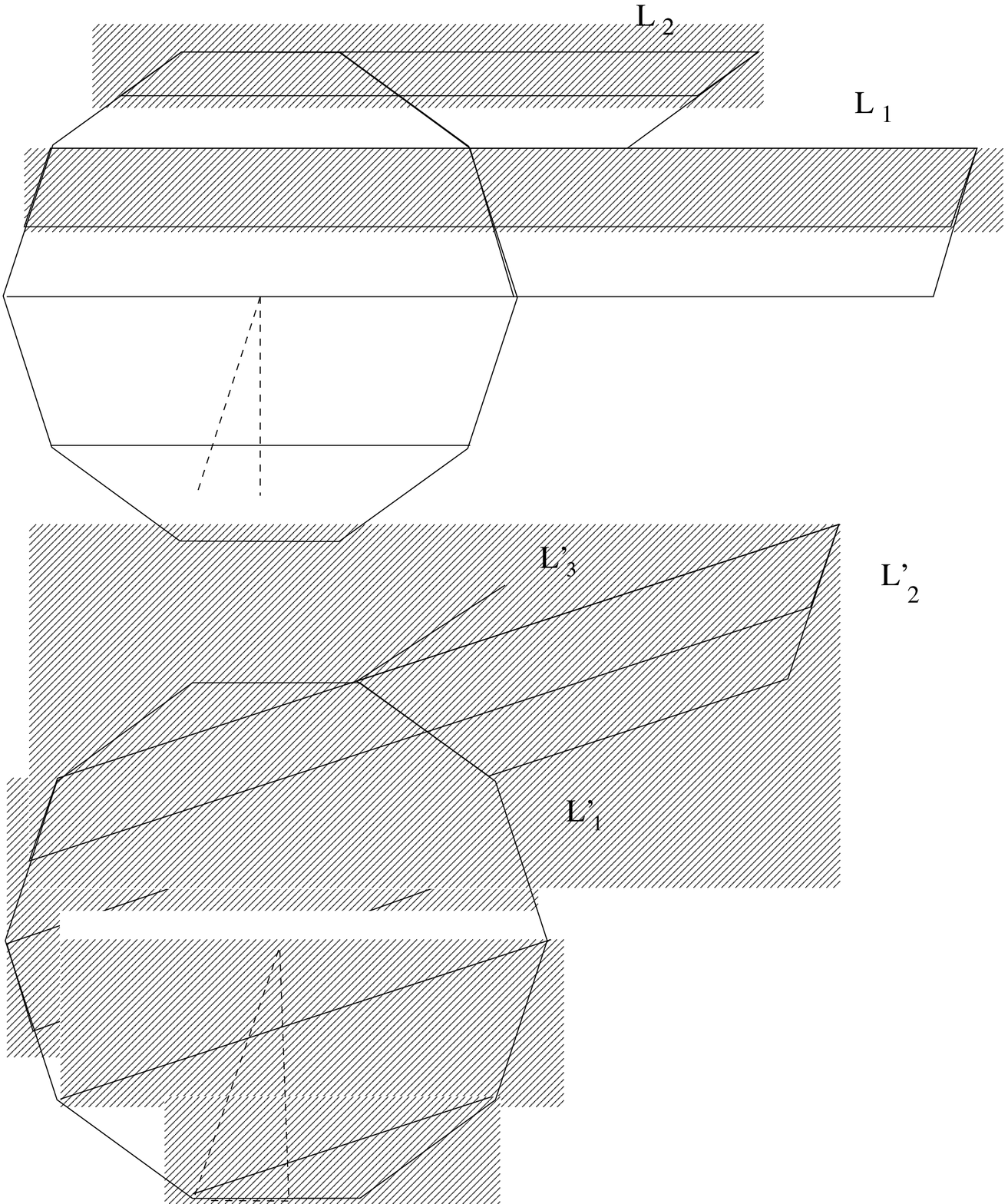,width=9cm}  
\end{center}
\caption{The elementary orbits in the decagon}
\label{octogone}
\end{figure}

For $n$ odd $(n=2 p +1)$, the lengths and widths are the following
\be
\label{longc}
L_{j}=4\sin\frac{2j\pi}{n}\cos\pn
\end{equation}
and
\be
\label{largc}
W_{j}=2 \sin\frac{2j\pi}{n}\sin\pn.
\end{equation}
with $1\leq j \leq p$.
\begin{figure}[ht]
\begin{center}
\epsfig{file=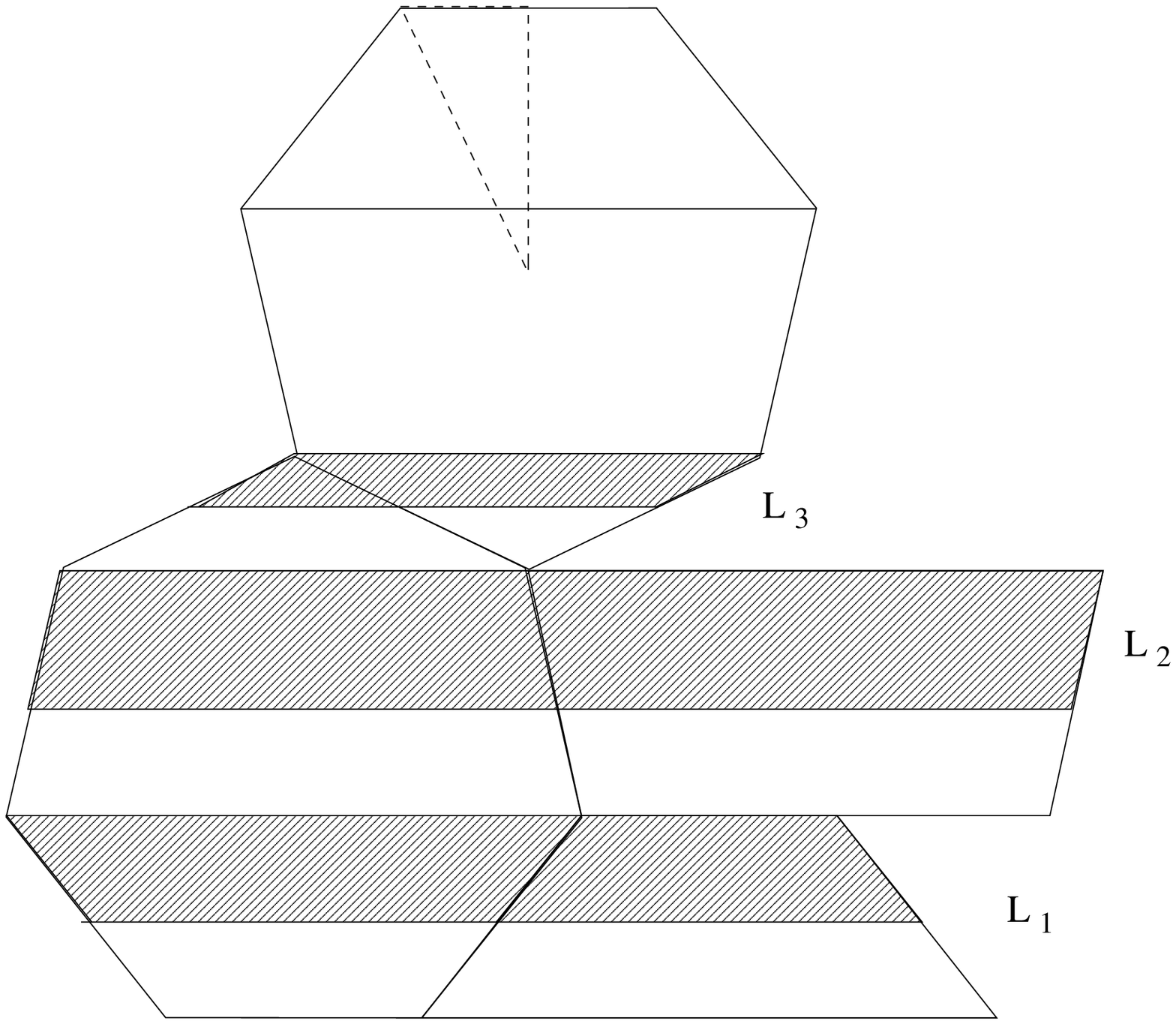,width=9cm}
\end{center}
\caption{The elementary orbits in the heptagon}
\label{pentagone}
\end{figure}
It is of interest to compute the ratio of the length of each periodic
orbit family to its width. From the above formulas it follows that for all
families, except the one with $j=1$ for even $n$, this ratio is the same
\begin{equation}
\frac{l}{w}=2\cot \frac{\pi}{n}.
\label{ratio}
\end{equation}
For the exceptional family (\ref{exceptional}) this ratio is 2 times smaller.

The unfolding of any family of periodic orbits gives an infinite strip 
of points of period  $L_p$ and width $W_p$. If
there is a group of invariance of the unfolded surface, it should include a
transformation which leaves invariant  periodic orbit strips. Assume that the
strip is oriented horizontally. In this case one  sees that the shift of
the form
\begin{equation}
\left( \begin{array}{cc} 1&\alpha \\0&1\end{array} \right)
\end{equation}
leaves points of the strip invariant provided that 
\begin{equation}
L_p=n \alpha W_p,
\end{equation}
where $n$ is an integer. 

Because for periodic orbits considered the ratio (\ref{ratio}) is constant
the invariance group should include the following transformation
\begin{equation}
\tn=\left(\begin{array}{cc}1 & 2\cot\pn \\
0&1\end{array}\right).
\label{taun}
\end{equation}
Veech proved \cite{veech}, \cite{vorobets} that the invariance group for
$\pi/n$ triangle is a discrete subgroup $\gn$ of $\sldr$ generated by the two 
elements (\ref{sigman}) and (\ref{taun}). Similarly to 
the relation (\ref{matorb}) for the square,
periodic orbits in this triangle are generated by the
action of $\gn$ over the elementary families of periodic orbits considered
above.

We shall call the members of these families ``basis orbits'' (and
label them by the index $i$). We define the
corresponding basis vectors $v_{i}$ by $v_{j}=(L_{j}, 0)$ and
$v'_{j}=(l'_{j}\cos(\pi/n), l'_{j}\sin(\pi/n))$, so that
$\{v_{i}\}=\{v_{j}, v'_{j}\}$ for $n$ even and $\{v_{i}\}=\{v_{j} \}$
for $n$ odd.

An element $g$ of the symmetry group $\gn $ has the following matrix
representation
\begin{equation}
g=\left (\begin{array}{cc}a&b\\c&d\end{array} \right ).
\end{equation}
The result of the action of this element to one of the basis vectors
$v_{i}=(v_{i1},v_{i2})$ gives the coordinates of a new primitive periodic 
orbit (more precisely, a periodic orbit situated on the boundary of 
periodic orbit pencil)
\begin{equation}
g\ v_i=\left (\begin{array}{cc}a&b\\c&d\end{array} \right )
\left (\begin{array}{c}v_{i1}\\v_{i2}\end{array} \right ),
\label{gv}  
\end{equation}
and the length of this primitive periodic orbit is the length of this vector.

The first family of basis vectors  for even $n$ and for all basis vectors
for odd $n$ can be chosen in the form  $v_i=(L_i,0)$ and the lengths of
periodic orbits generated by applying the group $\gn$ is 
\begin{equation}
L_g=\sqrt{a^2+c^2}L_i,
\label{lengths1}
\end{equation}
where $a$, $c$ are matrix elements of $g$ (\ref{gv}). The second basis
periodic orbits, $v_i'$,  are obtained from horizontal vectors by rotation
by $\pi/n$. But the matrix corresponding to the inverse of this  rotation
\begin{equation}
r=\left (\begin{array}{cc}\cos \frac{\pi}{n}&\sin \frac{\pi}{n}\\
-\sin \frac{\pi}{n}&\cos \frac{\pi}{n}\end{array} \right )
\end{equation}
does not belong to our group $\gn$. Nevertheless, this matrix plays the role
of a Hecke operator, namely, even if it does not belong to $\gn$ the
conjugation of any matrix from this group does belong to $\gn$: if
$g\in\gn$, then $r^{-1}gr \in\gn$.
To prove it let us note that 
\begin{equation}
r^{-1} \sigma_n^p r=\sigma_n^p
\end{equation}
where $\sigma_n$ is the generator (\ref{sigman}) because all
rotations commute, and it is easy to check that
\begin{equation}
r^{-1}\tau_n r=-\sigma_n \tau_n^{-1},\;r^{-1}\tau_n^{-1}=\tau_n \sigma_n^{-1}.
\end{equation}
The right-hand sides of these relations belong to $\gn$ and as  all
matrices from $\gn$ can be written as a product of generators we get
$r^{-1}gr \in\gn$ for $g\in\gn$.

Using this conjugation one can rotate the second family of periodic orbits
for even $n$ by $-\pi/n$ and the lengths of the orbits generated by the
vectors $v'_{i}$ will be related to
matrix elements of  $\gn$ by the same relation as in (\ref{lengths1})
\begin{equation}
L_g'=\sqrt{a^2+c^2}l_i'.
\label{lengths2}
\end{equation}
Therefore, to find the distribution of periodic orbit lengths it is
necessary to compute the distribution of $a^2+b^2$ for matrices from $\gn$,
which has been done for the modular group in the previous section.
Eq.~(\ref{calnveech}) can be derived the same way for $\gn$. 
According to the previous section one can compute the density of periodic
orbits and other quantities as well by investigation of the fundamental 
domain of $\gn$.

The distribution of areas of periodic orbit families is also easy to obtain:
as all matrices from $\gn $ have unit determinant,
the area covered by the pencil corresponding to $g\ v_{i}$ ($g\in\gn$) is the
same as the area covered by
the pencil corresponding to basis vectors $v_{i}$, i.e. it is equal to $L_{i} W_{i}$.
In other words, there is a one to one correspondence between pencils
of primitive periodic orbits and vectors $g\ v_{i}$ for $g\in\gn$. The discrete group
$\gn$ is related to periodic orbits in the $\pi/n$ right triangle in the same 
way as the modular group is related to periodic orbits in the square.

\subsubsection{The density of periodic orbits}\label{periodic}

The fundamental domains of the symmetry groups $\gn$ for $n$ even and  odd
are described  in Figs.~$\ref{domfondeven}$ and $\ref{domfondodd}$
respectively. For $n$ even, it is the union of two triangles with angles 
$2\pi/n, 0, 0\ $ on the Poincar\'e half-plane : the area of the domain is
$\vold=2\pi (n-2)/n$, and its width is $2\cot(\pi/n)$.

\begin{figure}[ht]
\begin{center}
\epsfig{file=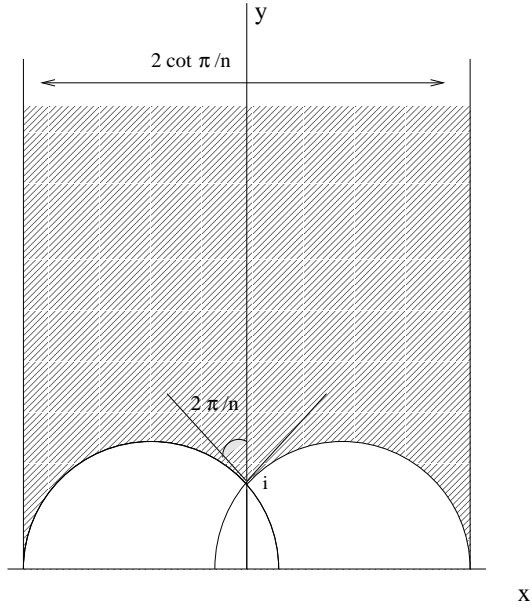,width=7cm}
\end{center}
\caption{The fundamental domain of $\gn$ for $n$ even}
\label{domfondeven}
\end{figure}

\begin{figure}[ht]
\begin{center}
\epsfig{file=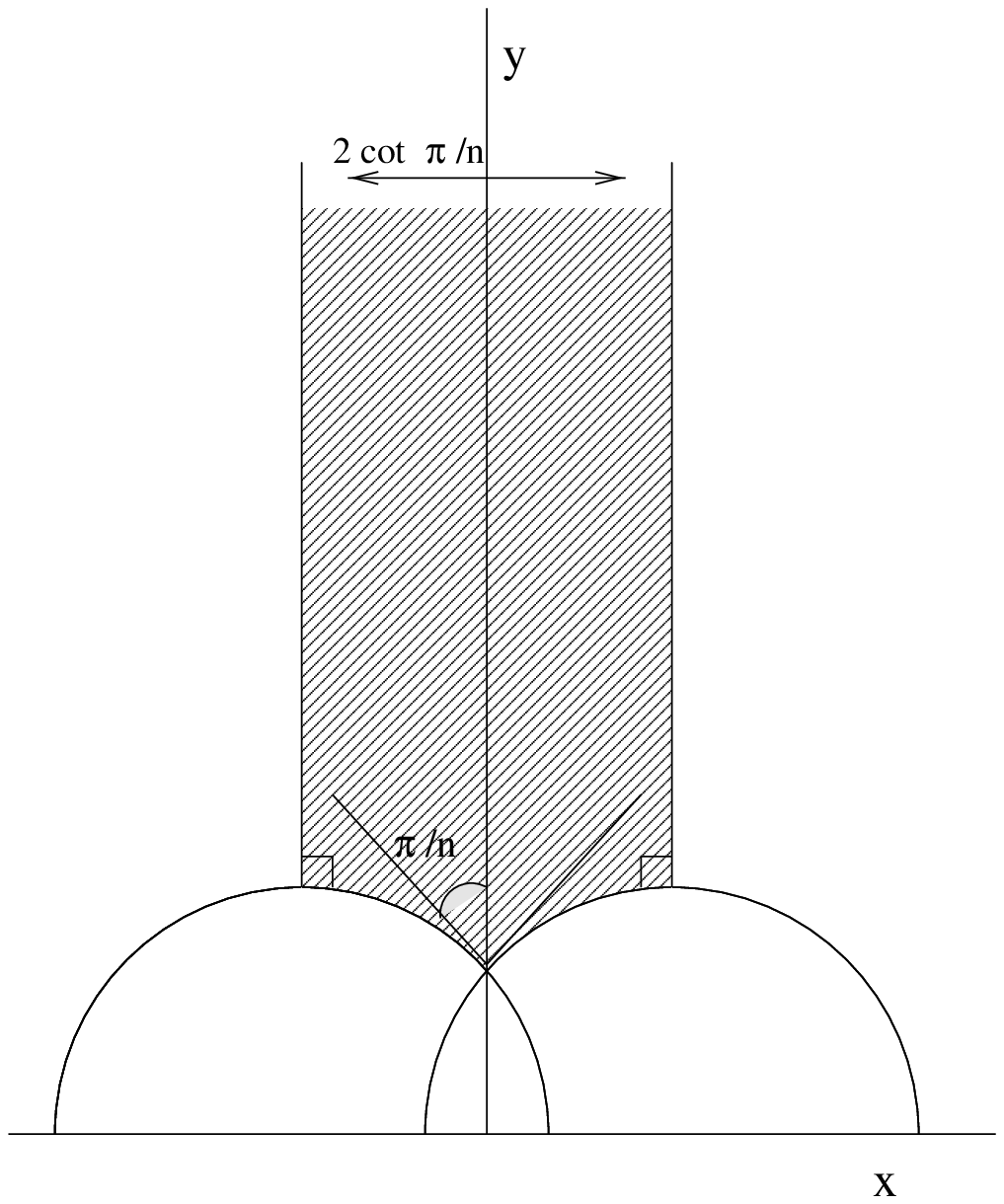,width=7cm}
\end{center}
\caption{The fundamental domain of  $\gn$ for $n$ odd}
\label{domfondodd}
\end{figure}

For $n$ odd  the two  triangles have angles $\pi/n$, $\pi/2$ and $0$,
therefore $\vold=\pi (n-2)/n$; the width of the fundamental domain is
$2\cot\pi/n$. These shapes of fundamental domains can
be obtained by taking into account that the group $\gn $ considered as a
group acting on the Poincar\'e upper-half plane as in Section~\ref{modular}
includes (i) the translation by $2\cot \pi/n$ and
(ii) the rotation around point $i$ by $2\pi/n$ for $n$ odd and by $4\pi/n$
for $n$ even.
This difference between even and odd $n$ is related to the fact
that  the rotation by angle $\pi$ corresponds to
the transformation $g\mapsto -g$, but these 2 matrices are represented by
the same function on $\calh$ (see $(\ref{gdz})$). For odd $n$ ($n=2q+1$),
the group generated by
the generator (\ref{sigman}) contains rotations by angles $2\pi j/(2q+1)$
for $j=0,1,\ldots 2q$. The value $j=q+1$ corresponds to the rotation by
$\pi+\pi/n$. As the rotation by $\pi$ is the identity transformation, the
rotation by $\pi/n$ belongs to $\gn$ and it is a primitive generator of the
subgroup $\{\sigma_{n}^{p}, p\in\zm\}$ of $\gn$.  

For even $n$ ($n=2q$), the rotation by $2\pi q/n$ is the identity, therefore
the rotation by $\pi/n$ does not belong to the group and the primitive
generator of the subgroup $\{\sigma_{n}^{p}, p\in\zm\}$ is the rotation by $2\pi/n$.

Due to $(\ref{calnveech})$, we now get the number of matrices 
\be
\label{gmnkl}
g=\left(\begin{array}{cc}a&b \\c&d\end{array}\right)\in \gn
\end{equation}
verifying $a^{2}+c^{2}\leq x^{2}\ $ :
\be
\caln (a^{2}+c^{2}\leq x^{2})=\left\{
\begin{array}{c} \displaystyle{\frac{n}{\pi (n-2)}}\cot\pn\ x^{2}\ \ \ \ \textrm{($n$ even)} \\
 \\
 \displaystyle{\frac{2 n}{\pi (n-2)}}\cot\pn\ x^{2}\ \ \ \ \textrm{($n$ odd)} 
\end{array}  
\right.
\end{equation}
These formulas give the total number of matrices verifying 
$a^{2}+c^{2}\leq x^{2}\ $. But due to the existence of rotation matrices
in the group $\gn$ each primitive periodic orbit length
appears a few times in the above calculations. This multiplicity corresponds
to different unfoldings of a given periodic orbit.

For $n$ odd the $2n$  matrices of the form
$\{\pm\beta_n^{k} g,\ 0\leq k\leq 2n-1\}$, where $\beta_{n}$ is the matrix of
rotation by $\pi/n$, give rise to one primitive periodic orbit. For $n$ even
there exist $n$ matrices of the form 
$\{\pm\sigma_n^{k} g,\ 0\leq k\leq n-1\}$ 
(where $\sigma_{n}$ is the matrix of rotation by $2\pi/n$ (\ref{sigman}))
which describe one periodic orbit.

For $g$ given by $(\ref{gmnkl})$, the length of
$g v_{i}$ is $L_{i} \sqrt{a^{2}+c^{2}}$. So the number of primitive
periodic orbits of type $g v_{i}$ less than $l$ is 
\be
\caln_{i, pp} (L_{p}<l)=
\frac{1}{\pi (n-2)}\cot\pn\ \left(\frac{l}{L_{i}}\right)^{2}.
\label{Ni}
\end{equation}

The number of all primitive periodic orbits is the sum over all such
contributions:
\begin{equation}
\label{calncala}  
\caln_{pp}(L_{p}<l)=\calc \frac{l^2}{\cala},
\end{equation}
where 
\begin{equation}
\cala =\frac{1}{4}\sin\frac{2\pi}{n}
\end{equation}
is the area of our triangle and
\begin{equation}
\calc =\frac{1}{2\pi (n-2)}\cos^2\pn \sum_{i}\frac{1}{L_{i}^{2}}.
\end{equation}
For $n$ odd ($n=2p+1$) Eq.~(\ref{longc}) gives
\begin{equation}
\sum_{k=1}^{p}\frac{1}{L_{k}^{2}}=\frac{1}{16\cos^2 \pi/n}
\sum_{k=1}^{p}\frac{1}{\sin^{2}(2\pi k/n)}.
\end{equation}
For $n$ even ($n=4p+2\epsilon$, $epsilon=0,1$) from Eqs.~(\ref{longa}), 
(\ref{longb}), and (\ref{exceptional}) it follows 
\begin{eqnarray}
&&\sum_{i=1}^{p}\frac{1}{L_{i}^{2}}=\frac{1}{16\cos^2 \pi/n}
\left (\sum_{j=1}^{p}\frac{1}{\cos^2(2j-1)\pi/n}+
\sum_{j=2}^{p+\epsilon}\frac{1}{\cos^2(2j-2)\pi/n}+4 \right )\nonumber\\
&&=
\frac{1}{16\cos^2 \pi/n}
\left (\sum_{k=1}^{(n-2)/2}\frac{1}{\cos^2 k\pi/n}+4 \right ).
\end{eqnarray}
The last sums can be calculated using the evident formulas (for another 
method of calculation see \cite{veech})
\begin{equation}
\frac{1}{\sin^2 x}=\sum_{q=-\infty}^{\infty} \frac{1}{(x-q\pi)^2},
\end{equation}
and 
\begin{equation}
\frac{1}{\cos^2\pi x/2}=\frac{4}{\pi^2}
\sum_{q=1}^{\infty}(\frac{1}{(2q-1-x)^2}+\frac{1}{(2q-1+x)^2}).
\end{equation}
Taking into account that $\sin^2(k\pi/n)=\sin^2((n-k)\pi/n)$ for odd $n$ and
performing the following transformations
\begin{equation}
\sum_{k=1}^{n-1}\sum_{q=-\infty}^{\infty}\frac{1}{(k-q n)^2}=
\sum_{t=-\infty}^{\infty}\frac{1}{t^2}-
\sum_{q=-\infty}^{\infty}\frac{1}{n^2q^2}=\frac{\pi^2}{3}(1-\frac{1}{n^2}),
\end{equation}
one obtains that for odd $n$
\begin{equation}
\sum_{k=1}^{(n-1)/2}\frac{1}{\sin^{2}(2\pi k/n)}=\frac{n^2-1}{6}.
\end{equation}
Similarly for even $n$
\begin{equation}
\sum_{k=1}^{(n-2)/2}\frac{1}{\cos^2 k\pi/n}=\frac{n^2-4}{6}.
\end{equation}
Therefore the value of constant $\calc$  is
\begin{equation}
\calc =\frac{1}{192 \pi (n-2)}\left \{
    \begin{array}{cc}(n^2-1)&\;\mbox{for odd}\;n\\ 
      (n^2+20)&\;\mbox{for even}\;n\end{array} \right . .
\label{asymptotic}
\end{equation}
This result corresponds to primitive periodic orbits with geometrically
different lengths: time-reversal orbits are not included in the summation.
They give an additional factor of 2 in Eq.~(\ref{asymptotic}).
In \cite{veech} the orbits corresponding to different unfoldings of the same
periodic orbit have been included in the asymptotic formula, which leads
asyptotically to the additional factor $n$ for even $n$ and $2n$ for odd $n$ in
Eq.~(\ref{asymptotic}). Furthermore, if one needs all periodic orbits including 
repetitions Eq.~(\ref{asymptotic})
should be multiplied by $\pi^2/6$ as in Section~\ref{square}. 

In Fig.~\ref{N8} we present numerical results of the cumulative density of
primitive periodic orbits in $\pi/8$ right triangular billiard (with area
$\cala=4\pi$) when all orbits (time-reversal and for different unfoldings)
are included. The solid line is the best quadratic fit to these data
\begin{equation}
\caln_{pp}(L_{p}<l)=.0294 l^2-.6055 l +.3617.
\end{equation}
One sees that this fit can hardy be distinguished from numerical results.
The theoretical prediction for the coefficient in front of $l^2$ is
$\calc/\cala$ according to  Eq.~(\ref{calncala}); $\calc$ is given by
Eq.~(\ref{asymptotic}), and has to be multiplied by $n$ since the numerical
computation has taken into account the repetitions of each orbit, and by $2$
as time-reversal orbits are taken into account as well:
\begin{equation}
 \frac{\calc}{\cala} =\frac{7}{24\pi^2}\approx .0295,
\end{equation}
which is in excellent agreement with numerical calculations.
\begin{figure}[ht]
\begin{center}
\begin{turn}{-90}  
\epsfig{file=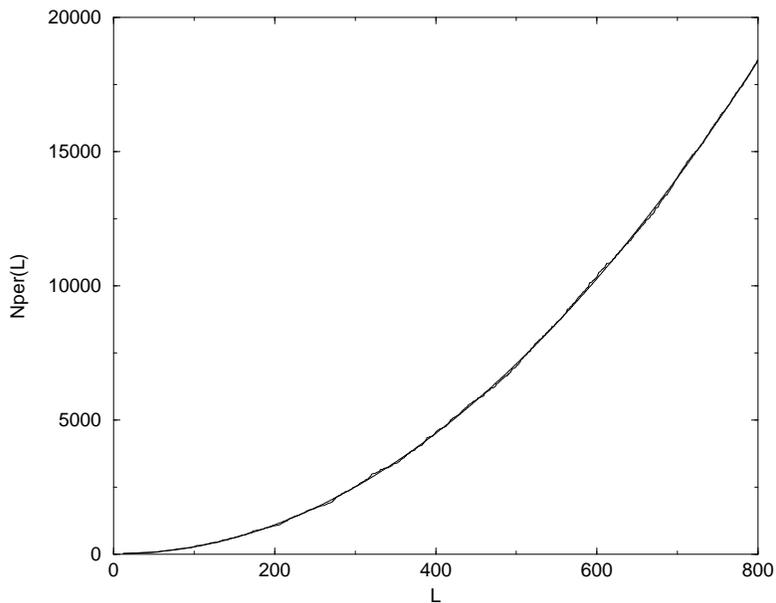,width=9cm}
\end{turn}
\end{center}
\caption{The cumulative density of periodic orbits for the $\pi/8$ right 
  triangular billiard.}
\label{N8}
\end{figure}

\subsection{Explicit calculation of the 2-point form factor for the  $\pi/n$
  right triangle}\label{calcul}

\subsubsection{First case: no degeneracy of the lengths}

We assume in this section that there is no degeneracy  between the
lengths of the periodic orbits (except the ones connected by the time-reversal
transformation) or more carefully, that there is no pair of primitive
periodic orbits whose 
lengths are commensurable. Since the lengths of the $g v_{i}$ are
proportional to the lengths of the $v_{i}$, the necessary requirement for
the validity of this condition is the absence of
commensurability relations between the $L_{i}$.

In this case the  2-point correlation form factor in the diagonal
approximation is  done by Eq.~(\ref{kperiodic}). The sum
over different periodic orbits can be split into a
sum over all types of periodic orbits, then the sum over periodic
orbits of each type can be replaced by an integral; since the density
$\rho_{i}$ of periodic orbits of type $i$ only takes into account once the
periodic orbit and its time-reverse, the degeneracy is $g_p=2$ and
\be
\sum_{p} g_{p}^{2}=4\sum_{i}\int_{0}^{\infty} d l \rho_{i}(l)\ .
\end{equation}
In  $(\ref{kperiodic})$, $\cala_{p}$ is the area occupied by a pencil of periodic orbits of length
$L_{p}$; but this area is the same for all trajectories belonging to the
same family $i$. So we just have to evaluate the area $\cala_{i}=L_{i} W_{i}$
occupied by an
elementary orbit of type $i$. The lengths $(\ref{longa})$, $(\ref{longb})$ and
$(\ref{longc})$ are
\be
\label{longueursa}
L_{k}=4\cos\frac{k\pi}{n}\cos\pn,\ \ \ 1\leq k \leq p-1
\end{equation}
and $L_{0}=2\cos\pi/n$ if $n=2 p$, and
\be
\label{longueursb}
L_{k}=4\sin\frac{2 k\pi}{n}\cos\pn,\ \ \ 1\leq k \leq p
\end{equation}
if $n=2p+1$.
The widths are only half the widths $W_{i}$ given by
$(\ref{larga})$, $(\ref{largb})$ and
$(\ref{largc})$ since each fundamental pencil is
symmetric with respect to the line joining two images of the $\pi/2$
corner of the triangle (see fig. $\ref{octogone}$ and $\ref{pentagone}$). So
the $W_{k}$ are
\be
\label{Largeursa}
W_{k}=\cos\frac{k\pi}{n}\sin\pn,\ \ \ 0\leq k \leq p-1
\end{equation}
if $n=2 p$, and
\be
\label{Largeursb}
W_{k}=\sin\frac{2 k\pi}{n}\sin\pn,\ \ \ 1\leq k \leq p
\end{equation}
if $n=2 p+1$. Then for small $\tau$
\be
\label{ktpetit}
\kt= 4\sum_{i}\frac{\cala_{i}^{2}}{8 \pi^{2}}\int_{0}^{\infty}\frac{1}{l \bar{d}}\
\delta\left( l -4\pi k \bar{d}\tau\right) \rho_{i}(l) d l,
\end{equation}
and replacing the density of orbits of type $i$ by its mean value 
\begin{equation}
\rho_i=\frac{\pi^{2}}{6}\frac{d \caln_{i, pp} (L_{p}<l)}{dl}
\label{rhoi}
\end{equation}
where $\caln_{i, pp} (L_{p}<l)$ is given by Eq.~(\ref{Ni}) we obtain,
when performing the integral,
\be
\kt=\frac{\cot \pi/n}{6\pi (n-2) \bar{d}}\sum_{k}W_{k}^{2}\ ;
\end{equation}
where the average density of states is $\bar{d}=\cala/4\pi$ with  
\be
\cala=\frac{1}{4}\sin\dpn
\end{equation}
is the area of the triangle.

The sum $(\ref{ktpetit})$ over the widths $(\ref{Largeursa})$ and
$(\ref{Largeursb})$ gives
\be
\sum_{k} W_{k}^{2}=\left\{
\begin{array}{c}
\frac{n+2}{4}\sin^{2}\pn,\ \ \ \ \ \textrm{$n$ even}\\ 
\\
\frac{n}{4}\sin^{2}\pn,\ \ \ \ \ \textrm{$n$ odd}
\end{array}
\right. .
\end{equation}
So we finally get
\be
\kt=\frac{n+\eps(n)}{3(n-2)}
\end{equation}
with
\begin{eqnarray}
\eps(n)=0 \;\;\textrm{when $n$ is odd}, \\
\eps(n)=2 \;\;\textrm{when $n$ is even}.
\end{eqnarray}

\subsubsection{Second case:  degeneracy of the lengths}
We have assumed in the previous Section that the lengths of all primitive
periodic orbits were non-commensurable. In the case of the  $\pi / n$ right triangle, there may exist a commensurability
relation between the $L_{k}$ given by $(\ref{longueursa})$ or $(\ref{longueursb})$
if there is one between the
$\cos (k\pi/n)$ ($0\leq k \leq p-1$) for $n$ even, or between the
$\sin (2 k\pi / n)$ ($1\leq k \leq p$) for $n$ odd.
It is shown in $\cite{jager}$ that if $n$ is an odd prime, there is no
such relation between the $\sin (2 k\pi / n)$. Ref. $(\cite{girstmair})$
deals with the case $(k,n)=1$ and gives the same conclusion. It seems that in
the general case, only one relation of that kind exists between the
$\cos (k\pi/n)$, which is 
\be
\label{dege}
2 \cos (\frac{n}{3}\frac{\pi}{n})=\cos(0)
\end{equation}
and that no relation exists between terms with sinus. Therefore 
the only degeneracy occurs in the case where $n$ is even and $3|n$, that
is $n\in 6\zm$. In that case, from $(\ref{densitepo})$ we get
\begin{eqnarray}
&&\left <d(E+\frac{\eps}{2})\ d(E-\frac{\eps}{2}) \right>= K^{diag}+ 
\nonumber \\  
&& 4\sum_{p^{+} \neq p^{'+}}
\frac{\cala_{p}\cala_{p'}}{16 \pi^{2}}
\frac{1}{2 \pi k \sqrt{l_{p} l_{p'}}} 
e^{i k(l_{p}-l_{p'})+i\frac{\eps}{4 k} (l_{p}+l_{p'})}
+\textrm{c.c.,}
\end{eqnarray}
where $K^{diag}$ is the usual diagonal approximation ($\ref{19})$.  
$p^{+}$ means that we only count for one orbit in the sum the orbit and its
time-reverse, therefore there is a coefficient $4$.

If there is a relation $m_{1} L_{1}=m_{2} L_{2}$
(with $m_{1}$ and $m_{2}$ coprime)
between two lengths of primitive periodic orbits,  we have a contribution
$\rdeg$ to the $2$-point correlation function $(\ref{corr})$ which
comes from orbits of lengths $q m_{1} L_{1}$ and $q m_{2} L_{2}$, 
$q\in\zm^{*}$ :
\begin{eqnarray}
\rdeg &  = & \frac{2\cala_{1}\cala_{2}}{4 \pi^{2} . 2 \pi
  k}\sum_{q}\frac{1}{\sqrt{q  m_{1} L_{1} q m_{2}  L_{2}}} e^{i\frac{\eps}{4
    k} (q  m_{1} L_{1}+ q m_{2} L_{2})} \nonumber \\
&=&  \frac{2\cala_{1}\cala_{2}}{4 \pi^{2} . 2 \pi k}\sum_{q}\frac{1}{q
  m_{1} L_{1}} e^{i\frac{\eps}{2 k}(q  m_{1} L_{1})}\ .
\end{eqnarray}
The sum over all repetition numbers $q$ of a function of $q L_{1}$ (where
$L_{1}$ is a primitive periodic orbit) can be replaced by an integral :
\be
\rdeg=\frac{\cala_{1}\cala_{2}}{4 \pi^{3} k}\int_{0}^{\infty} dl
\frac{1}{m_{1} l}e^{i\frac{\eps}{2 k} m_{1} l} \rho_{1}(l)
\end{equation}
where the density $\rho_{1}(l)$ is the density of periodic orbits of type
$1$ (that is, with length proportional to $L_{1}$) with
length less than $l$, given by $(\ref{rhoi})$. Performing the Fourier
transform $(\ref{defk})$ and the integral over $l$ gives
\be
K_{2}(0)=\frac{\cot \pi/n}{6\pi (n-2)\bar{d}}
 \left[\sum_{i}W_{i}^{2}+2\frac{W_{1} W_{2}}{m_{1}
  m_{2}}\right].
\end{equation}
In our case the degeneracy is given by $(\ref{dege})$ and
\be
W_{0} W_{\displaystyle{\frac{n}{3}}}=\frac{1}{2}\sin^{2}\pn.
\end{equation}
We finally obtain
\be
\label{ktaufinal}
K_{2}(0)=\frac{n+\eps(n)}{3(n-2)}
\end{equation}
with
\begin{equation}  
\eps(n)=\left \{
\begin{array}{ll}  
0 \ \ \ & \ \textrm{when $n$ is odd} \\
2 \ \ \ & \ \textrm{when $n$ is even and $3 \nmid n$}\\
6 \ \ \ & \ \textrm{when $n$ is even and $3 \mid n$}.
\end{array} \right ..
\end{equation}
This formula is the main result of our calculations for the triangular
billiards. It clearly demonstrated the peculiarities of spectral statistics
for pseudo-integrable systems. The non-zero value of the form factor ($<1$)
at the origin does not correspond to any random matrix ensemble but it
is typical for intermediate statistics \cite{gerland}, \cite{bgs}.

\subsection{Comparison with numerical calculations}\label{numerics}

To compare the prediction (\ref{ktaufinal}) with numerical results
we have computed 20000 levels for triangular billiards in the shape of a
right triangle with one angle $\pi/n$ for all $n=5,7\ldots,30$ 
(the case of $n=6$ is integrable). For each triangle we take levels from
15000 till 20000 and compute numerically the corresponding 2-point
correlation form factor. A typical result is presented in Fig.~\ref{k8}.
\begin{figure}[ht]
\begin{center}
\begin{turn}{-90}  
\epsfig{file=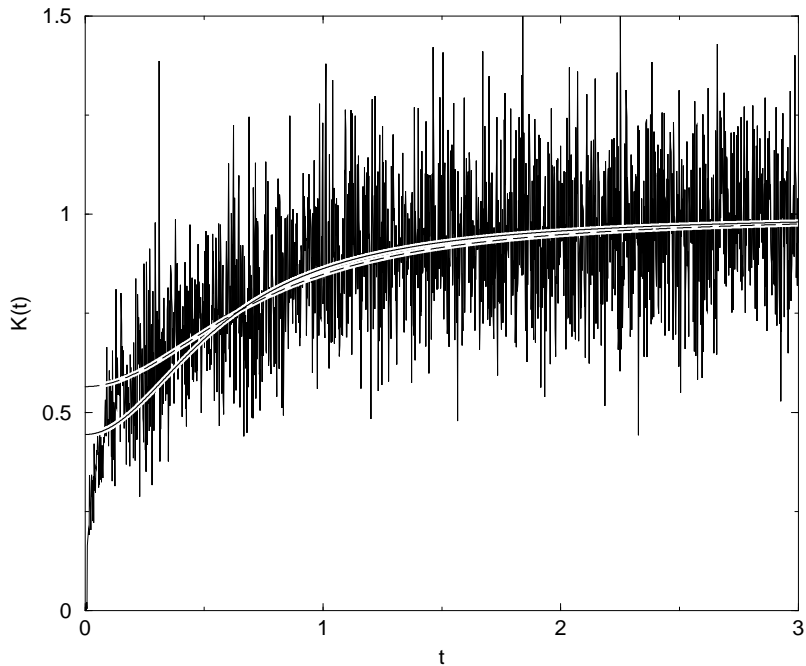,width=9cm}
\end{turn}
\end{center}
\caption{The 2-point form factor for $\pi/8$ right triangular billiard.}
\label{k8}
\end{figure}
From data like this, it is quite difficult to find the value of the form
factor at the origin because $\tau\to 0$ corresponds, according to
Eq.~(\ref{defk}), to an infinitely large energy difference in the $2$-point
correlation function: therefore numerically we always have $K_2(0)=0$. We
found it convenient first to fit the numerical data
to the following simple expression for the form factor, 
\begin{equation}
\label{kth}
K_2(\tau)=\frac{a^2-2 a +4 \pi^{2} \tau^{2}}{a^2 +4 \pi^{2} \tau^{2}}.
\end{equation}
and then from it compute $K_2(0)$.
\be
\label{kzero}
K_2(0)=1-\frac{2}{a}.
\end{equation}
The form (\ref{kth}) has been chosen because (i) one wants a simple expression, (ii)
when $\tau \rightarrow \infty$ the form factor should go to 1, (iii) to
describe the level repulsion it is necessary that
\begin{equation}
\label{intk}  
\int_0^{\infty} \left(1-K_2(\tau) \right) d\tau=\frac{1}{2},
\end{equation}
and (iv) the expression (\ref{kth}) when $a=4$ equals the form factor of the
so-called semi-Poisson model \cite{gerland}, \cite{bgs} which serves as a
reference point for intermediate statistics.

We stress that the above expression has no solid theoretical explanations and it
is used because it relatively well describes our numerical results. The only
fitting parameter is $K_2(0)$ related with $a$ by Eq.~(\ref{kzero}). We tried
two fitting procedures. First we fit Eq.~(\ref{kth}) for the data with all $\tau$
or, second, to decrease the influence of very small $\tau$, where numerical accuracy
is not very good, we did not consider the data with $0<\tau<0.25$. In
Fig.~\ref{k8} these two fits are presented. The first one gives
$K_2(0)\approx 0.44$ and the second one $K_2(0)\approx 0.565$. The expected value
$(\ref{ktaufinal})$ for $n=8$ is $5/9 \approx 0.56$.

In Fig.~\ref{kthfig} the results of such fitting procedures are given for
all triangles. Lower two curves correspond to these fits and the upper curve 
is the predictions $(\ref{ktaufinal})$. (Of course, only points are
important. Curves are presented for clarity.)
\begin{figure}[ht]
\begin{center}
\begin{turn}{-90}  
\epsfig{file=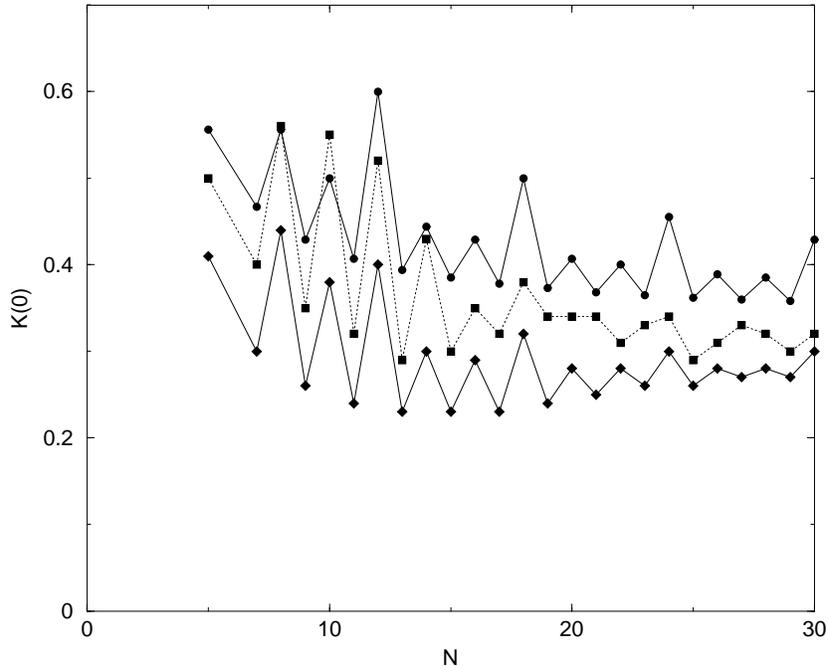,width=9cm}
\end{turn}
\end{center}
\caption{$K_2(0)$ for $\pi/n$ right triangles, $n=5$ to $30$. Circles are
  theoretical results (\ref{ktaufinal}). Squares are the fit (\ref{kth})
  when the region of small $\tau$, $0<\tau<.25$, is omitted. 
  Diamonds are the same fit but with all $\tau$.}
\label{kthfig}
\end{figure}
The numerical results quite well follow theoretical formula
$(\ref{ktaufinal})$ but there is a small shift which decreases when the
region of small $\tau$ is ignored.
This difference between the  curves seems to be a
consequence of the fact that the result $(\ref{ktaufinal})$ corresponds to
asymptotic limit $k\rightarrow \infty$ but numerical calculations have been
performed at large but finite energy.
\begin{figure}[ht]
\begin{center}
\begin{turn}{-90}  
\epsfig{file=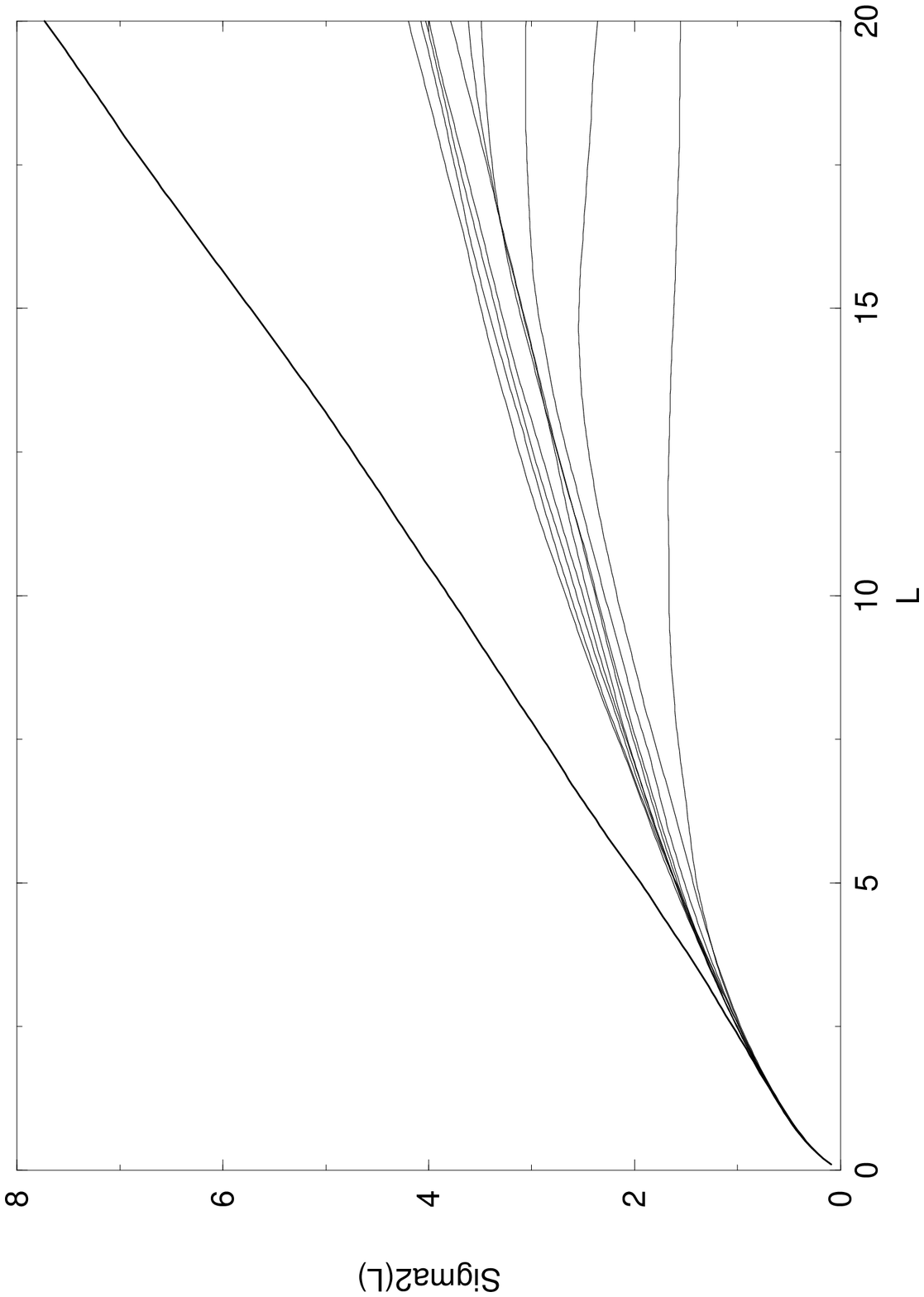,width=9cm}
\end{turn}
\end{center}
\caption{$\Sigma^{(2)}(L)$ for energy windows with higher and higher energy}
\label{pisurtrente}
\end{figure}
To check this point we present in  Fig.~\ref{pisurtrente}
the results of the calculation of the mean number variance,
$\Sigma^{(2)}(L)$, for the $\pi/30$ right triangle, in 10 energy intervals
$[8000 k, 8 000 (k+1)]$, $0\leq k \leq 9$ (the energy increases from bottom
to top). It is well known (see e.g. \cite{bohigas}) that the behavior
of the mean number variance at large distances is related with the
value of the form factor at the origin by simple formula
\begin{equation}
\Sigma^{(2)}(L)\rightarrow K_2(0) L \;\mbox{when }\;L\rightarrow \infty.
\end{equation}
From Fig.~\ref{pisurtrente} it is clearly seen that even for 80000 levels
the curve does not stabilize. To find its limiting behavior we extrapolate
point by point (with $L$ fixed) this ten curves with a fit $A(L)+B(L)/\sqrt{k}$.
(It means that for each $L$ we fit 10 points to find the best $A(L)$ and
$B(L)$.)
The limit curve (i.e. $A(L)$) is the most upper curve in Fig.~\ref{pisurtrente}.
It perfectly reproduces the expected
features of $\Sigma^{(2)}(L)$: it is a straight line with slope $K_2(0)=0.38$
corresponding to the expected value $(\ref{ktaufinal})$ for $n=30$.
In the same way, Prosen and Casati $\cite{proca}$ have computed
$\Sigma^{(2)}(L)$ for triangle billiards with angle $\pi /5$ for much larger
values of the energy, and it seems that such fit works well for their
calculations and the result  for $K_2(0)$ agrees with (\ref{ktaufinal}). These
(and other) calculations clearly demonstrate that the value of the 2-point
correlation form factor at the origin converges slowly to the theoretical
result (\ref{ktaufinal}) with increasing energy. This behavior may be a
consequence of the conjectured existence of two different terms
$(\ref{twoterms})$ in the  form factor and, in the final extent, a
manifestation of the strong diffraction in vicinity of optical boundaries.

\section{Rectangular billiard with a flux line}\label{flux}

\subsection{Preliminary calculations}
This section is devoted to the study of a rectangular billiard with
the Aha\-ro\-nov-Bohm flux line $\cite{bohmaharonov}$ at
a point $\vec{r_{0}}=(x_{0}, y_{0})$ inside the rectangle.
In the polar coordinates, $r$, $\varphi$, around this point the vector  
potential of the flux line has only $\varphi$ component 
\begin{equation}
A_{\varphi}=\displaystyle{\frac{\alpha}{ r}}
\label{fluxa}
\end{equation}
and the 2-dimensional Schr\"odinger equation for the motion in this 
potential is (when $\hbar =c=1$ and $m=1/2$) 
\begin{equation}
\left [ \frac{\partial^2}{\partial r^2}+
  \frac{1}{r}\frac{\partial}{\partial r}+
  \frac{1}{r^2} \left (\frac{\partial}{\partial \varphi}-i\alpha \right )^2
  +E_n \right ] \Psi_n(r,\varphi)=0.
\end{equation}
Similarly to triangular billiards discussed in previous Sections this model
belongs to the class of diffractive systems. The diffraction coefficient for
the scattering on the flux line 
(\ref{dflux}) diverges in the forward direction but as for pseudo-integrable
systems the contribution of diffractive orbits can be neglected when
computing the value of the 2-point correlation form factor at the origin.

It is well known that the Aharonov-Bohm potential (\ref{fluxa}) does 
not change classical trajectories
but gives an additional phase, $\Delta \phi$, when a trajectory 
turns $n$ times around the flux line
\be
\label{se}
\Delta \phi =2 n \pi \alpha.
\end{equation}
Therefore the contribution of periodic orbit to the trace formula
(\ref{densitepo}) will contain  an additional phase depending on 
the winding number of the trajectory around the flux line.

Periodic orbits in the rectangle of sides $a$, $b$ are determined by two 
integers $M$ and $N$ in the usual way and they are characterized by their length
\be
\label{longueurorbite}
l_{p}=\sqrt{(2 M a)^{2}+(2 N b)^{2}},
\end{equation}
the area occupied by the periodic orbit family, and the winding number around
the flux line. Each pencil of primitive periodic
orbits occupies an area $2 a b = 2\cala$, so its width is $2\cala / l_{p}$.
The images of the flux line in the unfolding of the rectangular billiard are
located at the points
\begin{equation}
((\zeta_{1} + 2 k) a, (\zeta_{2} + 2 k')b).
\end{equation}
Here $\zeta_{i}$ takes the values $\eps_{i}$ or $2-\eps_{i}$ ($i=1$, $2$),
where 
\begin{equation}
\eps_{1}=\frac{x_0}{a}\;\;\eps_{2}=\frac{y_0}{b}
\end{equation}
are the ratios of coordinates of the flux line to the corresponding sides,
and $k, k'\in\zm$ (see Fig.~\ref{unfoldedflux}).
\begin{figure}[ht]
\begin{center}
\epsfig{file=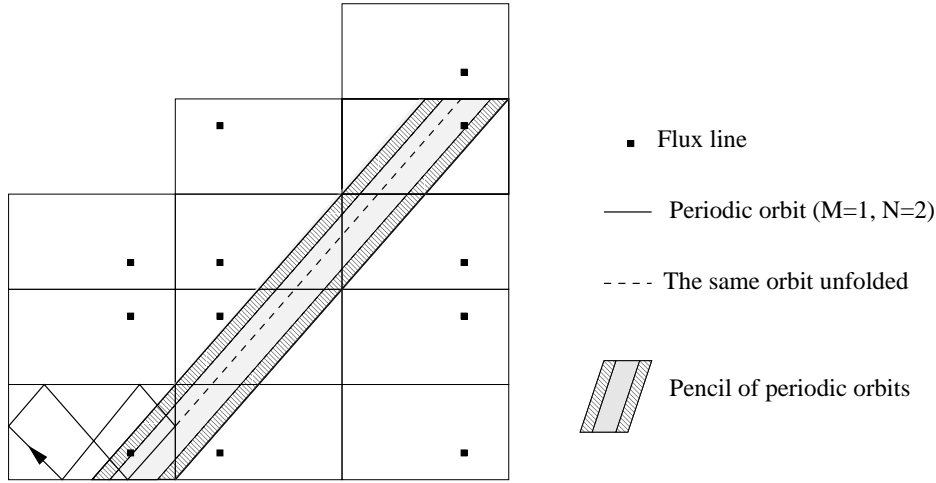,width=13cm}
\end{center}
\caption{An unfolded trajectory in the rectangle}
\label{unfoldedflux}
\end{figure}
Let us define $[x]$ as the largest integer less than or equal to $x$, 
so that
\be
[x] \leq x < [x]+1,
\end{equation}
and $\fx=x-[x] \in [0, 1[$. 

Each unfolded pencil of a primitive periodic orbit contains two and only two
images of the flux line, since it covers twice the area of the rectangle.
The periodic orbits  from this pencil parallel to the vector $(M,N)$ 
and going through the  images of the flux line (which we shall call the
saddle connections) 
split the pencil of  primitive periodic orbits parallel to  $(M,N)$ 
into three pencils of same length (see Fig.~\ref{unfoldedflux}). Only
the central strip is affected by the presence of the flux line
and any trajectory from this strip gets  a phase $ 2 \pi \alpha$ 
(according to $(\ref{se})$ and since the orbit is primitive). 
So the winding number of a periodic trajectory is nothing but the 
repetition number of a periodic orbit belonging to the central pencil.

To compute the widths of the central strip as a function of $M$ and $N$,
let us note that the algebraic distance from an image 
$((\zeta_{1} + 2 k) a, (\zeta_{2} + 2 k') b)$ of the flux line to 
the saddle-connection linking the points $(0,0)$ to $(2 M a, 2 N b)$ is
\be
\label{distsad}
d=\frac{2\cala}{l_{p}}\left(\zeta_{2}M-\zeta_{1}N+2 k' M - 2 k N \right)\ ;
\end{equation}
The two images of the flux line which are inside  the pencil are the two
nearest  points among the all images
$((\zeta_{1} + 2 k) a,(\zeta_{2} + 2 k') b)$ with integer $k$ and $k'$ that
have positive distance to the saddle-connection. 
They correspond to points such that the
distance $(\ref{distsad})$ is positive and less than $2\cala/l_{p}$, or less
than $1$ in units of $2\cala/l_{p}$. 

Let us set 
\begin{equation}
Q_{\pm}=[\epm],\;\; \eps_{\pm}=\{\epm\}.
\end{equation}
The four possible values of $(\zeta_{1},\zeta_{2})$
($(\eps_{1},\eps_{2})$, $(\eps_{1},2-\eps_{2})$, $(2-\eps_{1},\eps_{2})$, and
$(2-\eps_{1},2-\eps_{2})$) give four possible families for the distance 
(\ref{distsad}) :
\begin{eqnarray}
d_{i}&=&2 k_{i} \pm (Q_{+}+\eps_{+}) \nonumber\\
d_{i}&=&2 k_{i} \pm (Q_{-}+\eps_{-})
\end{eqnarray}
where $k_{i}$ is a certain integer which depends on $k, k', M$ and $N$.
Among these four families, exactly two points correspond to a distance
positive and less than $1$ : for instance
if $Q_{+}$ and $Q_{-}$ are even, only $2 k_{1}+ Q_{+}+\eps_{+}$ and
$2 k_{2} + Q_{-}+\eps_{-}$ can be
positive and less than one. So we must have $2 k_{1}+Q_{+}=0$ and
$2 k_{2}+Q_{-}=0$ and the two images of the flux line that are in the pencil
of primitive periodic orbits are at a distance
$\eps_{+}$ and $\eps_{-}$ from the saddle-connection $(0,0)-(2 M a, 2 N b)$ ;
if both $Q_{+}$ and $Q_{-}$ are odd,
the two distances are $1-\eps_{+}$ and $1-\eps_{-}$. Since the width of the
central strip is the difference between the two distances, it is in both
cases $|\eps_{1}-\eps_{2}|$. Dealing in the same way with the case where 
$Q_{+}$ and $Q_{-}$ have opposite parity,
we get that the width of the central strip in units of $2\cala /l_{p}$ is
\be
\eta=
\left\{
\begin{array}{l}
 \left| \eps_{-}-\eps_{+} \right|, \ \ \ \textrm{if $Q_{+}$ and $Q_{-}$ have
     the same parity} \\
\\
\left|1- \eps_{-}-\eps_{+} \right|, \ \ \ \textrm{if $Q_{+}$ and $Q_{-}$ have
     opposite parity}
\end{array}
\right. .
\end{equation}
Both cases can be summed up in the following formula :
\be
\label{etabig}
\eta=\varphi(x, y),
\end{equation}
with
\be
\label{varxy}
\begin{array}{l}
x=\emoins \\
y=\eplus 
\end{array},
\end{equation}
where
\be
\begin{array}{l}
\varphi(x, y)=\left| f(x) - f(y) \right|,
f(x)=(-1)^{[x]} \left( \{x\} - \dm \right) \\
\end{array}
\end{equation}
$f(x)$ is an even function of period $2$ ; if we restrict the study of $\varphi$
to  $[-1, 1]\times [-1, 1]$ we have
\be
\label{varphi}
\varphi (x,y)=
\left\{
\begin{array}{l}
|x+y| \ \ \  \textrm{if $x y \leq 0$}  \\
|x-y| \ \ \  \textrm{if $x y \geq 0$}
\end{array}
\right. ,
\end{equation}
and the Fourier expansion of $\varphi$ is
\be
\label{fourphi}
\varphi (x,y)=2 \sum_{n=1}^{\infty} \frac{\left( \cos\pi n x- \cos\pi n
  y\right)^{2}}{\pi^{2} n^{2}}.
\end{equation}
Using $(\ref{varxy})$ and $(\ref{fourphi})$, we obtain that the width of the
central strip associated to the orbit $(2 M,2 N)$ is the following
\be
\label{eta}
\eta=\frac{8}{\pi^2} \sum_{n=1}^{\infty} \frac{ \sin^{2}(\pi n \eps_{2} M)\sin^{2}(\pi n
  \eps_{1} N)}{n^{2}}.
\end{equation}

\subsection{Form factor for the billiard with flux line}
The value of the two-point correlation form factor in the diagonal
approximation  given by Eq.~(\ref{kperiodic}), when diffractive contributions 
have been neglected, still holds for  billiards with flux line. 
But $\cala_{p}$ now includes the
phase factor depending on the repetition number of the trajectory. The 
density of periodic orbits (\ref{densitepo}) becomes
\be
d_{p.o.}(E)=\frac{1}{2}\sum_{pp^{+},
  pp^{-}}\sum_{n=1}^{\infty}\frac{\cala_{p n}}{2\pi}\frac{1}{\sqrt{2\pi
k n l_{p}}} e^{i k n l_{p}-i\frac{\pi}{2}\nu_{p}-i\frac{\pi}{4}} + c.c.
\end{equation}
Here we distinguish between two types of orbits. The orbits associated with
a primitive orbit $pp^{+}$ have a phase $\exp (2 i \pi n\alpha)$  
for the orbit repeated $n$ times, and the total coefficient in the trace
formula associated with these orbits is
\be
\cala_{p^{+} n}=\cala_{p_{1}}+\cala_{p_{2}}e^{2 i \pi n\alpha}+\cala_{p_{3}},
\end{equation}
where $\cala_{p_{1}}$, $\cala_{p_{2}}$ and $\cala_{p_{3}}$ are the areas
covered respectively by the three strips in which the pencil of periodic
orbits splits. The  orbits associated with a primitive orbit
$pp^{-}$ have the complex conjugate  phase $\exp (-2 i \pi n\alpha)$ and
their contribution to the trace formula is proportional to
\be
\cala_{p^{-} n}=\cala_{p_{1}}+\cala_{p_{2}}e^{-2 i \pi n\alpha}+\cala_{p_{3}}.
\end{equation}
When the terms with same length $l_{p}=n l_{pp}$ are gathered together, we get
\be
d_{p.o.}(E)=\sum_{p^{+}}\frac{\cala_{n p}}{2\pi} \frac{1}{\sqrt{2\pi
k l_{p}}} e^{i k l_{p}-i\frac{\pi}{2}\nu_{p}-i\frac{\pi}{4}} + c.c.
\end{equation}
where 
\be
\cala_{n p}=\cala_{p_{1}}+\cala_{p_{2}}\cos(2 \pi n\alpha)+\cala_{p_{3}}
\end{equation}
and the sum $\sum_{p^{+}}$ goes over orbits $M, N \geq 0$. Equation
$(\ref{kz})$ now becomes
\be
\kt=\sum_{pp^{+}}\sum_{n=1}^{\infty}\frac{\cala_{n p}^{2}}{n^{2}}\frac{1}{2
  \pi^{2} l_{pp}\bar{d}}\
\delta\left( l_{pp}-\frac{4\pi k \bar{d}\tau}{n}\right).
\end{equation}
If $\eta_{p}$ is the width of the central strip expressed in units of
$2\cala / l_{p}$, we have
\begin{equation}
\cala_{n p}=2\cala (1-\eta_{p}+\eta_{p} \cos 2 \pi n \alpha)
=2\cala (1-2 \eta_{p} \sin^{2}\pi n \alpha),
\end{equation}
and (using the fact that $\bar{d}=\cala/4\pi$) the 2-point correlation
form factor at small $\tau$ is a sum of 3 terms
\begin{eqnarray}
\kt&=&\frac{8\cala}{\pi}\sum_{n=1}^{\infty}
\frac{1}{n^{2}}\sum_{pp}\frac{1}{l_{pp}}\delta\left(l_{pp}-\frac{\cala
k\tau}{n}\right)\nonumber\\
&-&\frac{32\cala}{\pi}\sum_{n=1}^{\infty}
\frac{\sin^{2}\pi n \alpha}{n^{2}}\sum_{pp}\frac{\eta_{pp}}{l_{pp}}\delta\left(l_{pp}-\frac{\cala
k\tau}{n}\right)\nonumber\\
&+&\frac{32\cala}{\pi}\sum_{n=1}^{\infty}\frac{\sin^{4}\pi n
  \alpha}{n^{2}}\sum_{pp}\frac{\eta_{pp}^{2}}{l_{pp}}\delta\left(l_{pp}-\frac{\cala
k\tau}{n}\right).
\label{3sommes}
\end{eqnarray}
The summation over primitive periodic orbits can be done by replacing the sum
by an integral, taking into account the density of primitive periodic
orbits. If $\eps_{1}$ and $\eps_{2}$ are rational numbers
\begin{equation}
\eps_{i}=\frac{p_{i}}{q_{i}},
\end{equation}
where $p_{i}$ and $q_{i}$ are co-prime integers, the width of the
central strip $(\ref{eta})$ only depends on the remainder $r_{1}$ of $M$ modulo
$q_{1}$ and $r_{2}$ of $N$ modulo $q_{2}$
\begin{equation}
\eta(r_1,r_2)=\frac{8}{\pi^2}
\sum_{n=1}^{\infty} 
\frac{1}{n^2} \sin^{2}(\pi \frac{p_1}{q_1} n r_1)\sin^{2}(\pi\frac{p_2}{q_2} n r_2).  
\label{etar}
\end{equation}
There are $q_{1} q_{2}$ periodic orbit families
\be
\label{familles}
M=q_{1} k + r_{1},\;\;N=q_{2} k'+r_{2},
\end{equation}
with $k, k'\in \nm$. 

To compute  sums in Eq.~(\ref{3sommes}) one needs to know the mean density of
primitive periodic orbits for each family, $\rho_{pp,r_1,r_2}(l)$.

Let $c$ be the greatest common divisor of $q_1$, $q_2$: $c=(q_{1}, q_{2})$.  If
$(r_{1}, r_{2}, c) \neq 1$, then $M$ and $N$ are not coprime
and there is no primitive periodic orbit. In the opposite case it is demonstrated in
Appendix A that
\begin{equation}
  \rho_{pp,r_1,r_2}(l)=\rho_{pp}(l)\alpha (r_1,r_2),
\end{equation}
where $\rho_{pp}(l)$ is the mean density for all primitive periodic orbits
in the rectangle (cf. Eq.~$(\ref{densiterect})$) with $M,N>0$
\begin{equation}
\rho_{pp}(l)=\frac{3l}{4\pi \cala},
\end{equation}
and 
\begin{equation}
 \alpha (r_1,r_2)=\frac{1}{q_1q_2 \prod_{p|\mbox{{\scriptsize lcf}}(q_1,q_2)}(1-1/p^2)}
\prod_{p|(q_1,r_1),p\nmid q_2}(1-\frac{1}{p})
\prod_{p|(q_2,r_2), p\nmid q_1}(1-\frac{1}{p}),
\end{equation}
and lcf$(q_1,q_2)$ is the least common factor of $q_1$, $q_2$. 

The knowledge of the mean density of periodic orbit families permits the
computation of mean values of different quantities depending on families. If
$f(r_1,r_2)$ is such a quantity its mean value is defined as follows
\begin{equation}
<f>=\sum_{\scriptstyle r_i\mod q_i \atop (r_1,r_2,c)=1}f(r_1,r_2)
\alpha (r_1,r_2).
\end{equation}
In particular 
\begin{eqnarray}
&&
\sum_{pp}\frac{\eta_{pp}^{\beta}}{l_{pp}}\ \delta\left(l_{pp}-\frac{\cala
k\tau}{n}\right) \label{etabeta}\\
&&=\sum_{{\scriptsize \begin{array}{c}r_i\mod q_i\\(r_1,r_2,c)=1\end{array}}}
\eta^{\beta} (r_1,r_2)\int_{0}^{\infty}\frac{1}{l}
\rho_{pp,r_1,r_2}(l)\delta\left(l-\frac{\cala k\tau}{n}\right) 
=\frac{3}{4 \pi \cala} <\eta^{\beta}>.\nonumber
\end{eqnarray}
The sums over $n$ that appear in  $(\ref{3sommes})$ can  be computed using
the standard formula
\be
\sum_{n=1}^{\infty}\frac{\cos 2\pi n x}{(2 \pi
  n)^{2}}=x^{2}-x+\frac{1}{6}, \ \ \ \ \textrm{for $0\leq x \leq 1$}.
\end{equation}
It gives
\be
\label{bernoulla}
\sum_{n=1}^{\infty}\frac{\sin^{2} \pi n \alpha
  }{n^{2}}=\frac{\pi^{2}}{2}\ba (1-\ba)
\end{equation}
and
\be
\label{bernoullb}
\sum_{n=1}^{\infty}\frac{\sin^{4} \pi n \alpha}{n^{2}}=\frac{\pi^{2}}{4}\ba
\end{equation}
where $\bar{\alpha}$ is the fractional part $\{\alpha\}$ of the flux through the
rectangle when $0\le \{\alpha\}\le 1/2$ and 
$\bar{\alpha}=1-\{\alpha\}$ when $1/2 \le \{\alpha\}\le 1$. 

Using $(\ref{etabeta})$, $(\ref{bernoulla})$ and $(\ref{bernoullb})$ one
concludes that the 2-point correlation form
factor for $\tau\to 0$ $(\ref{3sommes})$ is the following 
\be
\label{keta}
K_{2}(0)=1-12\ba (1-\ba)<\eta> +6 \ba <\eta^{2}>.
\end{equation}
To use this formula it is necessary to know the values of $<\eta>$ and
$<\eta^2>$. In the case where both $\eps_{1}$ and $\eps_{2}$ are irrational
non-commensurable quantities the fractional parts $\{n \eps_{2} M\}$ 
and $\{ n\eps_{1} N\}$ cover the whole interval $[-1,1]$ and
$\eta$ and $\eta^{2}$ can be computed   by integrating expression $(\ref{varphi})$
of $\eta$ over the square $[-1,1]\times [-1,1]$. Simple calculations show
that in this case
\be
<\eta >=\frac{1}{3},
\end{equation}
\be
<\eta^{2} >=\frac{1}{6}.
\end{equation}
Therefore when the coordinates  of the flux line are non-commensurable
with the corresponding sides
\be
K_{2}(0)=1-3\ba +4 \ba^{2}.
\end{equation}
In Appendix B, it is shown that for all rational $\eps_i$ 
\be
<\eta>=\frac{1}{3},
\end{equation}
like in the irrational case. 
The average $<\eta^{2}>$ is more difficult to compute
analytically: we have found an analytical expression  only when $q_{1}$ divides
$q_{2}$ (or similarly $q_{2}$ divides $q_{1}$).

%present an exact expression only when $q_{1}$ divides
%$q_{2}$ (or  $q_{2}$ divides $q_{1}$). If $q_{2}=q$, $q_{1}=p q$ and
%$q=p_{1}^{\alpha_{1}} p_{2}^{\alpha_{2}}\cdots p_{n}^{\alpha_{n}}$ is the
%prime factor decomposition of $q$, then the expression for  $<\eta^{2}>$ can
%be derived with almost the same technique as used in the appendix for the
%computation of $<\eta>$; it yields
%\be
%<\eta^{2}>=\frac{1}{6}\left(1-\frac{\lambda_{q}}{q^{4}} \left(1+\frac{2
%  \beta_{p,q}}{p^{2}} \right) (-1)^{n}
%\prod_{i=1}^{n}p_{i}^{2}\right)
%\end{equation}
%where $\lambda_{q}=0$ if $4|q$,  $\lambda_{q}=4/3$ if $2|q$ but $4\nmid q$,
%and  $\lambda_{q}=1$ if $q$ is odd. $\beta_{p,q}=1$ if one of the prime
%factors of $p$ is one of the prime factors $p_i$ ($1\leq i\leq n$) of $q$,
%and $\beta_{p,q}=0$ otherwise.

Though the general formula for $<\eta^2>$ is cumbersome, the computation of 
the $ <\eta^2>$ for rational $\eps_1$ and $\eps_2$ can
easily be done  numerically using Eqs.~$(\ref{varxy})$ and $(\ref{varphi})$.
For small denominators the values of $ <\eta^2>$ are given in
Table~\ref{etanumerique}.
\begin{table}
\begin{center}
\begin{tabular}{|c|c|c|c|c|c|c|c|c|} 
\hline
 $\begin{array}{ccc} && \\ &&q_{1}\\ &q_{2}& \end{array}$  & 2 & 3 & 4 & 5 & 6 & 7 & 8 \\
\hline
 &&&&&&& \\
   2                                         & 1/3  & 2/9  & 1/4   &  2/9 &   2/9 & 2/9   & 11/48 \\
  &&&&&&&   \\
   3                                         & 2/9  & 2/9  & 13/72
   &17/90   & 5/27  & 47/252  &107/576  \\
 &&&&&&&  \\
   4                                         & 1/4  & 13/72  & 1/6  &
   61/360  & 1/6  & 85/504  &1/6  \\
 &&&&&&&   \\
   5                                         &2/9   &17/90   &61/360
   &14/75   & 89/540  & 37/210  & 167/960 \\
 &&&&&&&   \\
   6                                         & 2/9  &5/27   &1/6
   &89/540   &4/27   & 35/216  &47/288  \\ 
 &&&&&&&   \\
   7                                         & 2/9  & 47/252  & 85/504
   &37/210   & 35/216  & 26/147  &85/504  \\ 
 &&&&&&&   \\
   8                                         &11/48   &107/576   & 1/6  &
   167/960  & 47/288  & 85/504  & 1/6 \\ 
 &&&&&&&   \\
	 \hline
\end{tabular}
\end{center}
\caption{Value of $<\eta^{2} >$ for a rational flux line}
\label{etanumerique}
\end{table}

To check the obtained formulas we have computed numerically 1500 first
energy levels for the rectangular billiard with sides $a=4$ and $b=\pi$ and the
flux line with coordinates (from low left corner) $x_{0}=5 a/9$ and $y_{0}=11 b /20$.
The typical picture of $K_2(t)$ is shown in Fig.~\ref{K7}.
\begin{figure}[ht]
\begin{center}
\epsfig{file=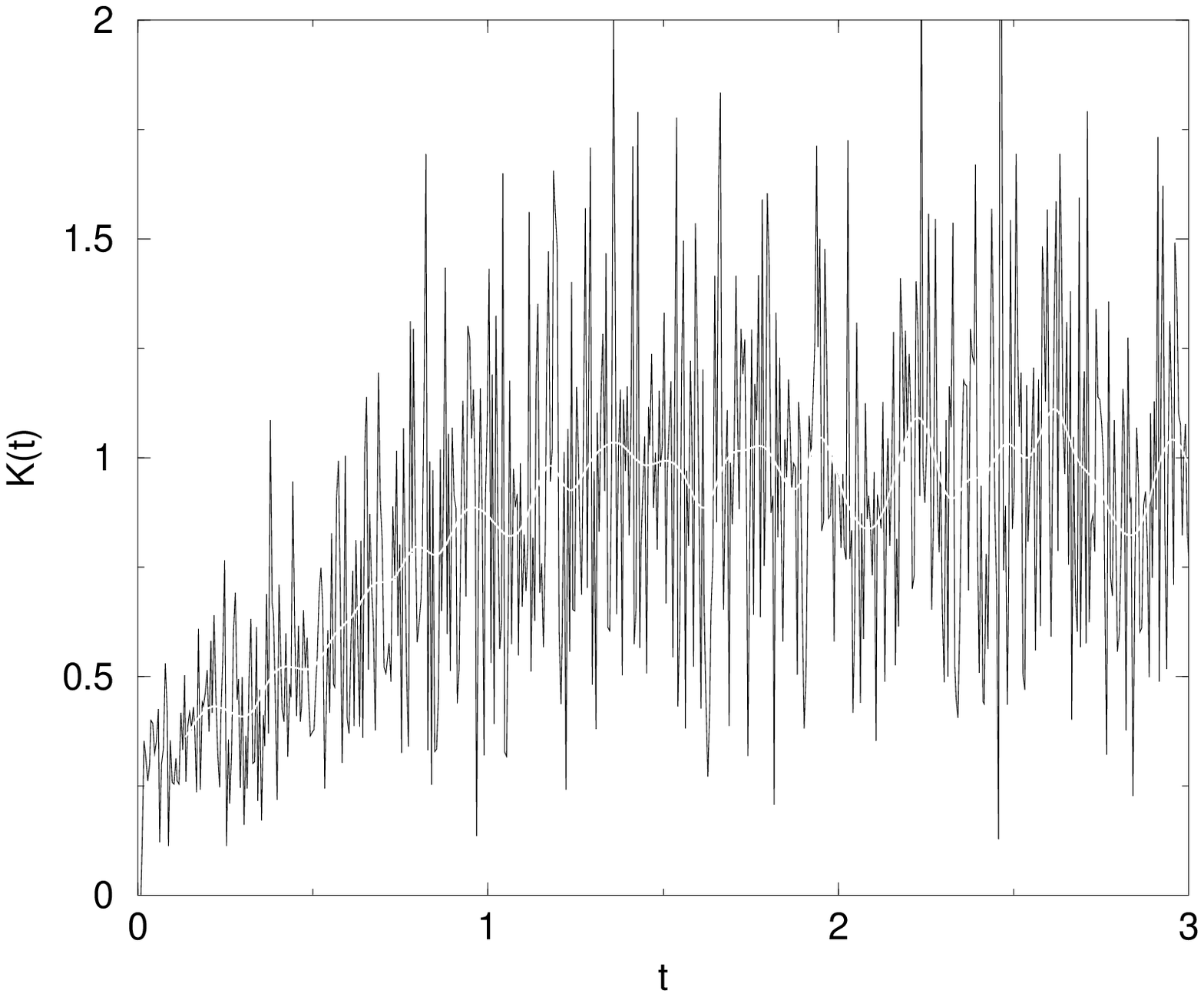,width=9cm}
\end{center}
\caption{The 2-point form factor for the rectangular billiard with flux
  line with $\alpha= 0.4$ and its smoothed value (white line).}
\label{K7}
\end{figure}
As for the triangular billiards discussed in previous Sections we
extrapolated $K_2(\tau)$ to small $\tau$ using the simple expression (\ref{kth}).
The results for different values of the flux are presented in Fig.~\ref{fluxalpha}.
\begin{figure}[ht]
\begin{center}
\begin{turn}{-90}  
\epsfig{file=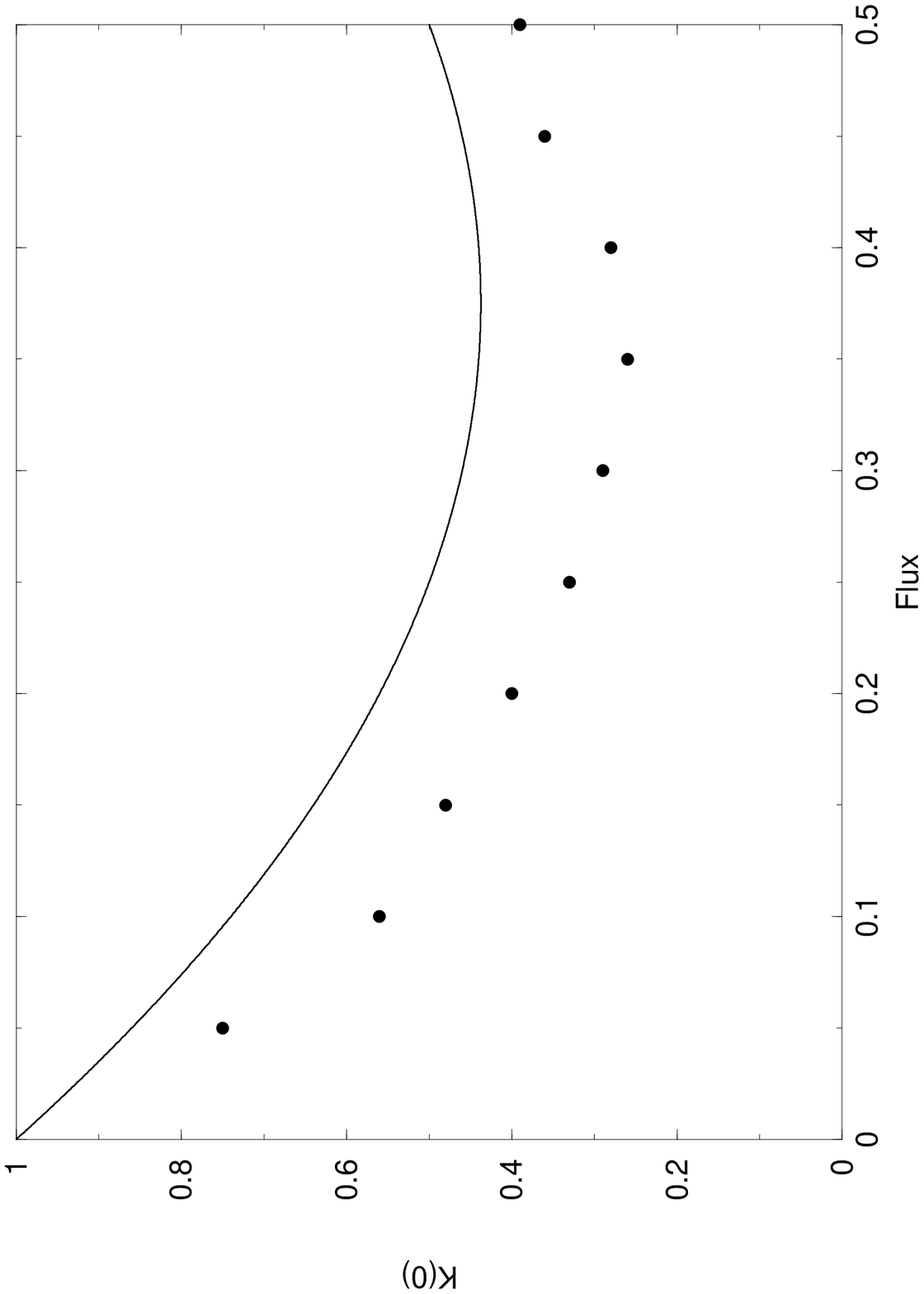,width=9cm}
\end{turn} 
\end{center}
\caption{$K_2(0)$ for different values of the flux $\alpha$ (points)  and
  the asymptotic theoretical prediction (solid line).}
\label{fluxalpha}
\end{figure}
We also check the following more suitable fit (which obeys the condition
(\ref{intk}) when $c=(1-b)^2$)
\begin{equation}
K_2(\tau)=\left \{\begin{array}{c} b+c\tau, \;\mbox{when }\ \tau<(1-b)/c\\
 1,\ \mbox{when }\ \tau>(1-b)/c\end{array}\right . .
\end{equation}
It gives practically the same results. The existing numerical precision does
not permit to distinguish these 2 fits.

In the case where $x_{0}=5 a/9$ and $y_{0}=11 b /20$, simple calculations
show that $<\eta^{2} >=4867/29160$; Eq.~$(\ref{keta})$ 
gives the expected value of $K_2(0)$
\be
K_{2}(0)=1-\frac{14573}{4860}\ba +4 \ba^{2}. 
\end{equation}
which corresponds to the solid curve in fig. $\ref{fluxalpha}$ (Note that the
coefficient of $\ba$ equals approximately $2.99$ and is practically indistinguishable from the
coefficient 3 for irrational $\eps_i$).

Similarly as for triangular billiards there is a small difference between the
theoretical and numerical curves. For triangular billiards where more
levels are available this difference slowly decreases with energy. We expect
the same behavior also for rectangular billiards with a flux line.

\section{Conclusion}\label{conclusion}

In this paper we have obtained  explicit expressions of  the 2-point correlation form 
factor $K_2(\tau)$ in the limit $\tau \rightarrow 0$ for a few typical examples 
of pseudo-integrable billiards : triangular billiards in the shape of 
right triangles with one angle equals $\pi/n$, and
rectangular billiards with a flux line. The obtained values of $K_2(0)$ differ from 
standard examples of spectral statistics (the Random matrix theory and the  Poisson
statistics), which confirm analytically  the peculiarities of spectral statistics
of  pseudo-integrable systems. The calculations have been performed by
analysing analytically the properties of classical periodic orbits of the
systems considered.  

In order to elucidate further the special properties of spectral statistics
of polygonal billiards, it would be of interest to compute $K_2(0)$ for  generic
triangular billiards without the Veech structure.
Moreover, we have taken into account only the diagonal terms and,
consequently, were able to obtain only $K_2(0)$. The computation of the next terms 
in the expansion of $K_2(\tau)$  in powers of $\tau$ should include the exact 
ressumation of singular contributions, coming from the diffraction close 
to the optical boundaries.  The solutions of these problems require the 
development of new methods beyond the ones used in this paper.

\section*{Appendix A}
\label{appendA}
Periodic orbits in a rectangle with sides $a$, $b$ are determined by 2
integers $M$ and $N$ which count the difference of coordinates of initial,
$(x_i,y_i)$, and final, $(x_f,y_f)$, points
\begin{equation}
x_f=x_i+2aM,\;\;y_f=y_i+2bN.
\end{equation}
The length of the periodic orbit is the geometrical length of this vector
\begin{equation}
L_p=\sqrt{(2aM)^2+(2bN)^2}.
\end{equation}
The mean cumulative density and the corresponding quantity for
primitive periodic orbits (when $M$, $N$ are co-prime integers) can be
computed as for the square billiard (see Eqs.~(\ref{no}) and
(\ref{densiterect})). When $l\rightarrow \infty$ and if only positive $M$ are
considered 
\begin{equation}
N(L_p<l)\rightarrow \frac{\pi l^2}{8ab},
\end{equation}
and 
\begin{equation}
N_{pp}(L_p<l)\rightarrow \frac{3 l^2}{4\pi ab}.
\end{equation}
The purpose of this Appendix is the computation of the mean cumulative
density of primitive periodic orbits for periodic orbit families when
\begin{equation}
M\equiv r_1 \;\mod q_1,\;\;N\equiv r_2 \;\mod q_2.
\end{equation}
The asymptotics of $N_{pp}(L_p<l)$ when $l\rightarrow \infty$ is related
with the behavior at small $x$  of the $\Theta$-function associated with
these periodic orbits  
\begin{equation}
\Theta(x)=\sum_{pp} e^{-xL_p^2}.
\end{equation}
If
\begin{equation}
\Theta(x)\rightarrow \frac{C}{x^{\gamma}}, \;\;\mbox{when }\;x\rightarrow 0,
\end{equation}
then
\begin{equation}
N_{pp}(L_p<l)\rightarrow \frac{C}{\gamma \Gamma (\gamma)}l^{2\gamma},
\;\;\mbox{when }l\rightarrow \infty.
\label{Npp}
\end{equation}
We are interested in the following $\Theta$-function
\begin{equation}
\Theta(x)=\sum^{\infty}_{M,N=-\infty}   e^{-x((2aM)^2+(2bN)^2)},
\label{thetax}
\end{equation}
where the summation is performed over all integers $M$, $N$ with the
following constraints
\begin{equation}
(M,N)=1,\;\;M\equiv r_1 \mod q_1,\;\;N\equiv r_2 \mod q_2.
\label{family}
\end{equation}
Note that both positive and negative values of $M$, $N$ are considered. When
only positive $M$ are taken into account the formulas below have 
asymptotically factor $1/2$.

To impose the restriction $M\equiv r \mod q$ it is convenient to introduce
the $\delta$-function
\begin{equation}
\delta_{t,q}=\left \{ \begin{array}{cc}1,&\;\mbox{if }t\equiv 0 \mod q\\
0,& \mbox{otherwise}\end{array}\right . .
\end{equation}
Its explicit form may be the following
\begin{equation}
\delta_{t,q}=\frac{1}{q}\sum_{k=0}^{q-1}e^{2\pi i kt/q}.
\label{deltat}
\end{equation}
As in Section ~\ref{square} the condition $(M,N)=1$ can be taken into
account by the inclusion-exclusion principle
\begin{equation}
\sum_{(M,N)=1}f(M,N)=\sum_{M,N=-\infty}^{\infty}
\sum_{t=1}^{\infty}f(Mt,Nt)\mu(t),
\end{equation}
where $\mu(t)$ is the M\"obius function equal  $(-1)^n$ if $t$ is a product
of $n$ distinct primes, 0 if $t$ contains a squared factor, and $\mu(1)=1$.

Combining all the necessary restrictions one finds the final expression
for the $\Theta$-function (\ref{thetax})
\begin{eqnarray}
\Theta(x)&=&\frac{1}{q_1q_2}\sum_{M,N=-\infty}^{\infty} 
\sum_{k_i\mod q_i} \sum_{t=1}^{\infty}
\mu(t)e^{-4xt^2(M^2a^2+N^2b^2)}
\nonumber\\
& &\times e^{2\pi i k_1(M t-r_1)/q_1+2\pi i k_2(N t-r_2)/q_2}.
\end{eqnarray}
Using the Poisson summation formula
\begin{equation}
\sum_{M=-\infty}^{\infty} e^{-xM^2+2\pi i y M}=\sqrt{\frac{\pi}{x}}
\sum_{M=-\infty}^{\infty} e^{-\pi^2(M+y)^2/x},
\end{equation}
one obtains that 
\begin{eqnarray}
\Theta(x)&=&\frac{\pi}{4ab x q_1q_2} 
\sum_{k_i\mod q_i} \sum_{t=1}^{\infty}
\frac{\mu(t)}{t^2}\ e^{-2\pi i k_1 r_1/q_1-2\pi i k_2 r_2/q_2}
\nonumber\\
&\times& \sum_{M=-\infty}^{\infty} e^{-\pi^2(M+k_1t/q_1)^2/(4 x t^{2} a^{2})}
\sum_{N=-\infty}^{\infty} e^{-\pi^2(N+k_2t/q_2)^2/(4 x t^{2} b^{2})}.
\end{eqnarray}
When $x\rightarrow 0$ the dominant contribution comes from terms with zero
exponent, i.e. from terms with
\begin{equation}
M+\frac{k_1t}{q_1}=0,\;\;\mbox{and}\; N+\frac{k_2t}{q_2}=0,
\end{equation}
or $k_it\equiv 0 \mod q_i$. The asymptotics of the $\Theta$-function is
therefore the following
\begin{equation}
\Theta(x)=\frac{\pi}{4abx}\frac{1}{q_1q_2}\sum_{k_i\mod q_i}
\sum_{t=1}^{\infty} \delta_{k_1 t,q_1}\delta_{k_2 t,q_2}\frac{\mu (t)}{t^2}
e^{-2\pi i k_1 r_1/q_1 -2\pi i k_2 r_2/q_2}.
\end{equation}
Using the  representation~(\ref{deltat}) for these $\delta$-functions and
performing the summation over $k_i$ one gets that when $x\rightarrow 0$
\begin{equation}
\Theta(x)=\frac{\pi}{4abx}F(r_1,r_2),
\end{equation}
where
\begin{equation}
F(r_1,r_2)=\frac{1}{q_1q_2}\sum_{l_i\mod q_i}
\sum_{t=1}^{\infty} \delta_{l_1 t-r_1,q_1}\delta_{l_2 t-r_2,q_2}
\frac{\mu (t)}{t^2}.
\label{F}
\end{equation}
From Eq.~(\ref{Npp})  one concludes that the asymptotics of the mean cumulative
density of primitive periodic family (\ref{family}) (with $M,N>0$, i.e. with
a factor $1/4$) is 
\begin{equation}
N_{pp}(L_p<l)=\frac{\pi l^2}{16 ab}F(r_1,r_2).
\end{equation}
To perform the summation over $t$ in Eq.~(\ref{F}) it is necessary to know
the number of solutions of two equations
\begin{equation}
l_1t\equiv r_1 \mod q_1,\;\;l_2t\equiv r_2 \mod q_2.
\label{equations}
\end{equation}
It is well known (and can be easily checked) that the number of solutions of
the equation $ax\equiv b \mod q$ depends on the greatest common divisor of $a$ and
$q$, $(a,q)=d$. If $d=1$ there is 1 solution, $x\equiv ba^{-1} \mod q$. If
$d>1$ and $d \nmid b$ there is no solutions. If $d|b$ there is one solution,
$x_0=(b/d)(a/d)^{-1} \mod (q/d)$ and consequently, there are $d$ solutions
modulo $q$: $x_j=x_0+(q/d)j$, $j=0,\ldots d-1$. Therefore
\begin{eqnarray}
\label{fintermediaire}  
F(r_1,r_2)&=&\frac{1}{q_1q_2}\sum_{t=1}^{\infty}\frac{\mu (t)}{t^2}
(q_1,t) (q_2,t)  \delta_{(q_1,t),r_1} \delta_{(q_2,t),r_2} \nonumber\\
&=&\frac{1}{q_1q_2}\sum_{\scriptstyle d_1|(q_1,r_1)\atop d_2|(q_2,r_2)}
\sum_{\scriptstyle (q_1,t)=d_1 \atop (q_2,t)=d_2}\frac{\mu (t)}{t^2}\ d_1 d_2.
\end{eqnarray}
Terms corresponding to $(d_1,d_2)>1$ give a $0$ contribution to the sum,
since in that case $q_1, q_2,d_1$ and $d_2$ have a common factor, which
contradicts condition $(M,N)=1$. The sum $(\ref{fintermediaire})$ can
therefore be restricted to  $(d_1,d_2)=1$, and the sum over $t$ is now a sum
over $t'$ where $t=d_1 d_2 t'$.
Let us denote by $P$  the product of the prime factors of $q_{1}$ that do not
divide $c$ and by  $P'$ the product of the prime factors of $q_{2}$
that do not divide $c$. Now 
\begin{equation}    
(c,P)=(c,P')=(P,P')=1.
\label{PP}
\end{equation}
If $p$ is a prime dividing $d_1$ and $c=(q_1,q_2)$, then $p$ divides
$d_1=(q_1,t)$. Since it divides $c$ it also divides $q_2$, so $p|(q_2,t)$
and $(d_1,d_2)\neq 1$, which is impossible. So the prime divisors of $d_1$
have to be taken among the divisors of $P$, and in the same way  the prime
divisors of $d_2$ have to be taken among the divisors of $P'$.  Similarly, we
can check that if $p$ is a prime divisor of $c$ which divides $t'$, $p$
divides $d_1=(q_1,t)$, so $d_1$ would contain a prime factor of $c$, which
is impossible; and if $p$ is a prime factor of $P$ or $P'$, $p$ divides both
$d_1$ and $d_2$. So the sum over $t'$ must be restricted to $t'$ which do
not contain any prime divisor of $q_1$ or $q_2$. As $\mu(ab)=\mu(a)\mu(b)$
for co-prime $a$ and $b$, one gets
\be
F(r_1,r_2)=\frac{1}{q_1q_2}\sum_{d_1,d_2,t'}d_1d_2\frac{\mu(d_1)\mu(d_2)\mu(t')}{d_1^2d_2^2{t'}^{2}}
\end{equation}
where the sum is taken over all $d_1,d_2,t'$ verifying $ d_1|(P,r_1)$,
$d_2|(P',r_2)$, $t'\nmid q_1$, $t'\nmid q_2$. Using the identity
\begin{equation}
\prod_{p|k}(1-\frac{1}{p^{s}})=\sum_{\delta|k}\frac{\mu(\delta)}{\delta^{s}},
\end{equation}
we get
\begin{eqnarray}
F(r_1,r_2)&=&\frac{1}{q_1q_2}\prod_{p\nmid q_1,p\nmid q_2}(1-\frac{1}{p^2})
\prod_{p|(P, r_1)}(1-\frac{1}{p})
\prod_{p|(P',r_2)}(1-\frac{1}{p})\nonumber \\
&=&\prod_{\mbox{{\scriptsize all }}p}(1-\frac{1}{p^2})\alpha (r_1,r_2)=\frac{6}{\pi^2}\alpha (r_1,r_2),
\end{eqnarray}
where 
\begin{equation}
\label{alphar1r2}
\alpha (r_1,r_2)=\frac{1}{q_1q_2 \prod_{p|\mbox{{\scriptsize lcf}}(q_1,q_2)}(1-1/p^2)}
\prod_{p|(q_1,r_1), p\nmid c}(1-\frac{1}{p})
\prod_{p|(q_2,r_2), p\nmid c}(1-\frac{1}{p}),
\end{equation}
and lcf$(q_1,q_2)$ is the least common factor of $q_1$, $q_2$
(lcf$(q_1,q_2)=q_1 q_2/c$).

It is also instructive to check directly that $\alpha (r_1,r_2)$ are normalized
correctly
\begin{equation}
\sum_{\scriptstyle r_i\mod q_i \atop (r_1,r_2,c)=1}\alpha
(r_1,r_2)=1,
\label{198}
\end{equation}
where as above $c=(q_1,q_2)$. We use once more the inclusion-exclusion principle 
\begin{equation}
\label{inclexcl}  
\sum_{\scriptstyle r_i\mod q_i \atop (r_1,r_2,c)=1}f(r_1,r_2)=
\sum_{t|c}\mu(t) \sum_{\scriptstyle r_i\mod q_i\atop t|r_i }f(r_1,r_2).
\end{equation}
If we set
\begin{equation}
D=q_1q_2\prod_{p| \mbox{{\scriptsize lcf} }(q_1,q_2)}(1-\frac{1}{p^2}),
\end{equation}
we get, with $(\ref{alphar1r2})$ and $(\ref{inclexcl})$,
\begin{eqnarray}
\sum_{\scriptstyle r_i\mod q_i\atop (r_1,r_2,c)=1}\alpha(r_1,r_2)
&=&\frac{1}{D}\sum_{t|c}\mu(t)\sum_{r_1 ( t|r_{1})}\ \sum_{r_2 (t|r_{2})}\ 
\sum_{\delta_{1}|(P, r_{1})}\frac{\mu(\delta_{1})}{\delta_{1}}
\sum_{\delta_{2}|(P', r_{2})}\frac{\mu(\delta_{2})}{\delta_{2}}\nonumber\\
&=&\frac{1}{D}\sum_{t|c}\mu(t)\sum_{\delta_{1}|P}\ \sum_{\delta_{2}|P'}\  
\sum_{r_1 (t\delta_{1}|r_1)}\ \sum_{r_2 (t\delta_{2}|r_2)}
\frac{\mu(\delta_{1})}{\delta_{1}} \frac{\mu(\delta_{2})}{\delta_{2}} \nonumber\\
&=&\frac{1}{D}q_{1} q_{2} \sum_{t|c}\frac{\mu(t)}{t^{2}}\sum_{\delta_{1}|P}
\frac{\mu(\delta_{1})}{\delta_{1}^{2}}\sum_{\delta_{2}|P'} 
\frac{\mu(\delta_{2})}{\delta_{2}^{2}}.
\label{sommalpha}
\end{eqnarray}
Here we have used the fact that if $t|c$ and $\delta |P$, $t$ and $\delta$ have
no common factor. In the above sums it is always understood that the
summation over $r_{i}$ goes only from $r_i=0$ to $q_i-1$.
But the last sum in Eq.~(\ref{sommalpha}) exactly equals $D$ because 
$c P P'=\mbox{lcf}(q_1,q_2)$ and Eq.~(\ref{198}) holds.

\section*{Appendix B}
\label{appendB}
In the same way one can compute the mean value of $\eta$ defined in Eq.~(\ref{etar}) 
\begin{eqnarray}
<\eta>=\sum_{\scriptstyle r_i\mod q_i\atop (r_1,r_2,c)=1}   
\eta (r_1,r_2)\alpha (r_1,r_2)=\frac{8}{\pi^2 D}
\sum_{n=1}^{\infty}\frac{1}{n^{2}}
\sum_{t|c}\mu(t) \\
\sum_{r_1 (t|r_{1})}\ \sum_{r_2 (t|r_{2})}\sum_{\delta_{1}|(P, r_{1})}\sum_{\delta_{2}|(P', r_{2})}
\frac{\mu(\delta_{1})}{\delta_{1}}\frac{\mu(\delta_{2})}{\delta_{2}}
 \sin^{2}\left(\pi n \frac{r_{1}}{q_{1}}\right) 
 \sin^{2}\left(\pi n \frac{r_{2}}{q_{2}}\right).\nonumber
\end{eqnarray}
Since
\be
\sum_{t\delta|r}\sin^{2}\pi n \frac{r}{q} =
\frac{q}{2 t\delta}\left( 1- \delta_{n t \delta,q}\right),
\end{equation}
one obtains 
\begin{eqnarray}
\label{eqeta}
<\eta>&=&\frac{2 q_{1} q_{2}}{\pi^2 D}
\sum_{t|c}\frac{\mu(t)}{t^{2}}\sum_{\delta_{1}|P}
\frac{\mu(\delta_{1})}{\delta_{1}^{2}}\sum_{\delta_{2}|P'} 
\frac{\mu(\delta_{2})}{\delta_{2}^{2}} \nonumber\\
&&\sum_{n=1}^{\infty}\frac{1}{n^{2}}\left( 1- \delta_{n t\delta_{1},q_1}\right)
\left( 1- \delta_{n t \delta_{2},q_2}\right).
\label{204}
\end{eqnarray}
The sum over $n$ includes 4 terms. The first is the sum over all $n$
\begin{equation}
\sum_{n=1}^{\infty}\frac{1}{n^{2}}=\frac{\pi^2}{6}.
\label{first}
\end{equation}
The second sum has the restriction that $n=(q_1/t\delta_1)m$ and   
\be
\sum_{n=1}^{\infty}\frac{1}{n^{2}}\delta_{n t \delta_{1},q_1}=
\frac{\pi^2}{6}\left(\frac{t\delta_{1}}{q_{1}}\right)^{2}.
\end{equation}
The third sum is the same but with the substitution
$1\rightarrow 2$. The fourth sum incorporates two restrictions,
$nt\delta_1 \equiv 0 \mod q_1$ and $nt\delta_2 \equiv 0 \mod q_2$.
Remembering the definition of $P$ and $P'$ (see (\ref{PP})) one concludes
that in this last case the restriction is $n=(cPP'/(t\delta_1\delta_2))m$ and 
\be
\sum_{n=1}^{\infty}\frac{1}{n^{2}}\delta_{n t\delta_{1},q_1}
  \delta_{n t\delta_{2},q_2}
=\frac{\pi^2}{6}\left(\frac{t\delta_{1}\delta_2}{cPP'}\right)^{2}.
\end{equation}
Performing the summation over $\delta_i$ and $t$ in Eq.~(\ref{204}) one notes that
all three last sums will have as a factor 
\be
\label{sommed}
\sum_{\delta_{1} |P}\mu(\delta_{1})\ \ \ \ {\textrm or}\ \ \ \ 
\sum_{\delta_{2} |P}\mu(\delta_{2}).
\end{equation}
But  for any $K\geq 2$ we have
\be
\sum_{\delta |K}\mu(\delta)=0.
\end{equation}
Since $q_{1}$ and $q_{2}$ are greater than $1$, it is impossible that
simultaneously $c=P=1$, or $c=P'=1$, the terms $(\ref{sommed})$ equal zero. 
Therefore only the term (\ref{first}) survives and $(\ref{eqeta})$ gives
\be
<\eta>=\frac{1}{3D} q_{1} q_{2}\sum_{t|c}\frac{\mu(t)}{t^{2}}\sum_{\delta_{1}|P}
\frac{\mu(\delta_{1})}{\delta_{1}^{2}}\sum_{\delta_{2}|P'} 
\frac{\mu(\delta_{2})}{\delta_{2}^{2}}.
\end{equation}
These sums are exactly equal to $D$ and finally we get
\be
<\eta>=\frac{1}{3}.
\end{equation}


\begin{thebibliography}{99}

\bibitem{bohigas} Bohigas, O.: Chaos and Quantum Mechanics, Giannoni, M.-J.,
  Voros, A. and Zinn-Justin, J. eds., Les Houches Summer School Lectures LII,
  1989 (North Holland, Amsterdam, 1991), p. 87.
\bibitem{mehta}M. L. Mehta, {\it Random Matrix Theory} (Springer, New York,
  1990).
\bibitem{bohigasgiannoni} Bohigas, O., Giannoni, M.-J., Schmit, C.:
  Characterization of chaotic quantum spectra and universality of level
  fluctuation laws. Phys.
  Rev. Lett. {\bf 52}, 1 (1984).
\bibitem{berrytaborc} Berry, M. V., Tabor, M.: Level clustering in the
  regular spectrum. Proc. Roy. Soc. Lond. {\bf
    356}, 375 (1977). 
\bibitem{andreev} Andreev, A. V., Altshuler, B.L.: Spectral Statistics
  beyond Random Matrix Theory. Phys. Rev. Lett. {\bf
    75}, 902 (1995); Agam, O., Altshuler, B.L., Andreev, A.V.: Spectral
  Statistics from disordered to chaotic systems. Phys. Rev. Lett {\bf 75}, 4389 (1995). 
\bibitem{keating} Bogomolny, E. B., Keating, J.P.: Gutzwiller's Trace
  formula and Spectral Statistics: beyond the diagonal approximation. Phys. Rev. Lett. {\bf
    77}, 1472 (1996). 
\bibitem{marklof} Marklof, J.: Spectral Form Factors of Rectangle Billiards. 
  Commun. Math. Phys. {\bf 199} 169 (1998).
\bibitem{bb} Balian, R., Bloch, C.: Distribution of eigenfrequencies for the
  wave equation in a finite domain: Eigenfrequency density oscillations. Ann. Phys. (N.Y.) {\bf 69}, 76 (1972).
\bibitem{Gutz} Gutzwiller, M. C.: Chaos and Quantum Mechanics,  Giannoni, M.-J.,
  Voros, A. and Zinn-Justin, J. eds., Les Houches Summer School Lectures LII,
  1989 (North Holland, Amsterdam, 1991), p. 201.
\bibitem{berrytabor}Berry, M. V., Tabor, M.: Closed orbits and the regular
  bound spectrum. Proc. Roy. Soc. Lond. {\bf
    349}, 101 (1976), {\it ibid}  J. Phys. A: Math. Gen. 
  {\bf  10}, 371 (1977).
\bibitem{berryrichens} Richens, P. J., Berry, M. V.: Pseudointegrable
  systems in classical and quantum mechanics. Physica D {\bf 2}, 495
  (1981).  
\bibitem{shimizu} Shudo, A., Shimizu, Y.: Extensive numerical study of
  spectral statistics for rational and irrational polygonal billiards. Phys. Rev. E {\bf 47}, 54 (1993).
\bibitem{ugerland} Bogomolny, E. B., Gerland, U., Schmit, C.: Models of
  intermediate spectral statistics. Phys. Rev. E {\bf
    59}, 1315 (1999).
\bibitem{shklovskii} Schklovskii, B.I. {\it et al.}: Statistics of spectra
  of disordered systems near the metal-insulator transition. Phys. Rev. B {\bf 47},
  11487 (1993).
\bibitem{bogomolny} Bogomolny, E. B., Pavloff, N., Schmit, C.: Diffractive
  corrections in the trace formula for polygonal billiards. Phys. Rev. E
  {|bf 61}, 3689 (2000).
\bibitem{berry} Berry, M. V.: Semiclassical theory of spectral rigidity. Proc. Roy. Soc. A {\bf 400}, 229 (1985).
\bibitem{veech} Veech, W. A.: Teichm\"uller curves in moduli space,
  Eisenstein series and an application to triangular billiards. Invent. Math. {\bf 97} (1989), 553-583.
\bibitem{vorobets} Vorobets, Y. B.: Planar structures and billiards in
  rational polygons: the Veech alternative. Russian Math. Surveys {\bf
    51}, 5 (1996), 779-817.
\bibitem{keller} Keller, J.P.: Geometrical theory of diffraction. J. Opt. Soc. Am. {\bf 52}, 116 (1962).
\bibitem{rosenquist}Vattay, G., Wirzba, A., Rosenqvist, P.E.: Periodic orbit
  theory of diffraction. Phys. Rev.
  Lett {\bf 73}, 2304 (1994).
\bibitem{pavloff} Pavloff, N., Schmit, C.: Diffractive orbits in quantum
  billiards. Phys. Rev. Lett {\bf 75}, 61
  (1995); {\bf 75}, 3779 (E) (1995). 
\bibitem{bohmaharonov} Aharonov, Y., Bohm, D.: Significance of
  electromagnetic potentials in the quantum theory. Phys. Rev. {\bf 115}, 485
  (1959).
\bibitem{bogomolny2} Bogomolny, E. B.: Action correlations in integrable
  systems. Nonlinearity {\bf 13}, 947 (2000).
\bibitem{varenna} Bogomolny, E. B., in New Directions in Quantum Chaos.  
  Proc. of the International School of Physics ``Enrico Fermi'', course
  CXLIII, eds. Casati, G., Guarneri, I., Smilansky, U. 333 (1999).  
\bibitem{gerland} Bogomolny, E. B., Gerland, U., Schmit, C.: Singular
    statistics, Phys. Rev. E (2000), to be published.
\bibitem{courburneg} Bogomolny, E. B., in  Quantum Dynamics of Simple
  Systems,
  Oppo, G.L., Barnett, S.M., Riis, E., Wilkinson, M. eds., The Forty Fourth
  Scottish Universities Summer School in Physics (Institute of physics
  publishing, Bristol and Philadelphia, August 1994), p. 17.
\bibitem{masurun} Masur, H.: Holomorphic Functions and Moduli, Vol. I
  (Berkeley, CA, 1986), Math. Sci. Research Inst. Publ. 10, Springer-Verlag,
  New York - Berlin 1988, pp. 215-228.
\bibitem{masurdeux} Masur, H.: The growth rate of trajectories of a
  quadratic differential. Ergod. Theory of Dynam. Syst. {\bf 10} (1990), 151-176.
\bibitem{terras} Terras, A.: Harmonic analysis on symmetric spaces and
    applications, I, Springer-Verlag New York - Berlin 1985, p. 206.
\bibitem{jager} Jager, H., Lenstra jr., H. W.: Linear independance of
  cosecant values. Nieuw Archief Wisk. (3) {\bf 23} (1975), 131-144.
\bibitem{girstmair} Girstmair, K.: Character coordinates and annihilators of
  cyclotomic numbers. Manuscripta Math. {\bf 59} (1987), 375-389.
\bibitem{bgs} Bogomolny, E. B., Gerland, U., Schmit, C.: Short-range plasma
  model for intermediate spectral statistics, Europ.  Phys. J. B (2000), 
  to be published.
\bibitem{proca} Casati, G., Prosen, T.: Quantum chaos in triangular billiards,
  unpublished (2000).

\end{thebibliography}
\end{document}